\documentclass[12pt]{article}

\usepackage{scicite}

\usepackage{times}
\usepackage[utf8]{inputenc}
\usepackage[T1]{fontenc}
\usepackage{amsmath,amsfonts,amssymb,braket}
\usepackage{bm}
\usepackage{cleveref}
\usepackage{graphicx}
\usepackage{upgreek}
\usepackage{color}
\usepackage{scicite}

\usepackage[version=4]{mhchem}
\usepackage{siunitx}

\newenvironment{sciabstract}{%
\begin{quote} \bf}
{\end{quote}}

\title{Observation of the Magnonic Dicke Superradiant Phase Transition} 

\author
{Dasom Kim$^{1,2,3\dagger}$, Sohail Dasgupta$^{4,\dagger}$, Xiaoxuan Ma$^{5,\dagger}$, Joong-Mok Park$^{3}$,\\ Hao-Tian Wei$^{4}$, Liang Luo$^{3}$, Jacques Doumani$^{1,2}$, Xinwei Li$^{6}$, Wanting Yang$^{5}$,\\ Di Cheng$^{3,7}$, Richard H. J. Kim$^{3}$, Henry O. Everitt$^{2,8,9}$, Shojiro Kimura$^{10}$,\\ Hiroyuki Nojiri$^{10}$, Jigang Wang$^{3,7}$, Shixun Cao$^{5,\ast}$, Motoaki Bamba$^{11}$,\\ Kaden R. A. Hazzard$^{4,8,12}$,  Junichiro Kono$^{2,4,8,13,\ast}$\\
\\
\normalsize{$^{1}$Applied Physics Graduate Program, Smalley--Curl Institute, Rice University,}\\
\normalsize{Houston, TX 77005, USA}\\
\normalsize{$^{2}$Department of Electrical and Computer Engineering, Rice University,}\\
\normalsize{Houston, TX 77005, USA}\\
\normalsize{${^3}$Ames National Laboratory, Ames, IA 50011, USA}\\
\normalsize{$^{4}$Department of Physics and Astronomy, Rice University, Houston, TX 77005, USA}\\
\normalsize{$^{5}$Department of Physics, International Center of Quantum and Molecular Structures,}\\
\normalsize{and Materials Genome Institute, Shanghai University, Shanghai, 200444, China}\\
\normalsize{$^{6}$Department of Physics, National University of Singapore, 117551, Singapore}\\
\normalsize{$^{7}$Department of Physics and Astronomy, Iowa State University, Ames, IA 50011, USA}\\
\normalsize{$^{8}$Smalley–Curl Institute, Rice University, Houston, TX 77005, USA}\\
\normalsize{$^{9}$DEVCOM Army Research Laboratory-South, Houston, TX 77005, USA}\\
\normalsize{$^{10}$Institute for Materials Research, Tohoku University, Sendai 980-8577, Japan}\\
\normalsize{$^{11}$Department of Physics, Yokohama National University, Yokohama 240-8501, Japan}\\
\normalsize{$^{12}$Department of Physics and Astronomy, University of California, Davis, CA 95616, USA}\\
\normalsize{$^{13}$Department of Materials Science and NanoEngineering, Rice University,}\\
\normalsize{Houston, TX 77005, USA}\\
\\
\normalsize{$^\ast$To whom correspondence should be addressed; E-mail: sxcao@shu.edu.cn, kono@rice.edu}\\
\normalsize{$^\dagger$These authors contributed equally.}
}

\date{}

\begin{document} 


\baselineskip24pt


\maketitle 

\newpage


\begin{sciabstract} 
Two-level atoms coupled with single-mode cavity photons are predicted to exhibit a quantum phase transition when the coupling strength exceeds a critical value, entering a phase in which atomic polarization and photonic field are finite even at zero temperature and without external driving. However, this phenomenon, the superradiant phase transition (SRPT), is forbidden by a no-go theorem due to the existence of the diamagnetic term in the Hamiltonian.  Here, we present spectroscopic evidence for a magnonic SRPT in ErFeO$_3$, where the role of the photonic mode (two-level atoms) in the photonic SRPT is played by an Fe$^{3+}$ magnon mode (Er$^{3+}$ spins). The absence of the diamagnetic term in the Fe$^{3+}$--Er$^{3+}$ exchange coupling ensures that the no-go theorem does not apply. Terahertz and gigahertz magnetospectroscopy experiments revealed the signatures of the SRPT -- a kink and a softening, respectively, of two spin--magnon hybridized modes at the critical point.

\end{sciabstract}


\section*{Main text}
\subsection*{Introduction}
An ensemble of two-level atoms can exhibit coherence through cooperative interaction with a single-mode quantized radiation field. Such cooperative optical processes have been extensively studied since the pioneering work of Dicke in the context of superradiance~\cite{Dicke1954} and have recently attracted much-renewed interest in cavity quantum electrodynamics (QED)~\cite{Garraway2011, FornDiaz2019, FriskKockum2019}, condensed matter physics~\cite{Cong2016, peraca2020, Schlawin2022}, and quantum information science~\cite{Buzek2005}. In contrast to superradiance phenomena, recent cavity QED studies of materials have focused on thermal equilibrium modified by cavity-enhanced vacuum electromagnetic fields; see Fig.\,1A (top panel). 

A profound consequence of the Dicke model is a quantum phase transition, the superradiant phase transition (SRPT)~\cite{HEPP1973360}, where when the strength of the cooperative light--mater coupling ($g$) exceeds a critical value, a static coherent electric or magnetic field and a finite atomic polarization appearing spontaneously and simultaneously. Realization of the SRPT has been of much interest, but
its occurrence in thermal equilibrium has been a subject of debate~\cite{Bamba2016a, Stokes2019}. A no-go theorem exists~\cite{Rzafmmodeotzlseziewski1975}, while various methods have been proposed to circumvent the no-go theorem~\cite{Nataf2010, Bamba2016b, Andolina2019, Nataf2019, Garziano2020, Bamba2022}.
%
At the core of the no-go theorem is the diamagnetic term (also known as the $A^2$ term) that inevitably appears in the minimal-coupling Hamiltonian describing the electric-dipole light--matter interaction; this term adds positive energy to the system, causing the ground state to be robust against the SRPT~\cite{Rzafmmodeotzlseziewski1975, Li2018NP}. Figure\,1B plots the frequencies of the upper and lower polaritons ($\omega_+$ and $\omega_-$, respectively, normalized by $\omega_\text{0}$) as a function of $g/\omega_\text{0}$, where $\omega_\text{0}$ is the cavity frequency, in the presence (dashed line) and absence (solid line) of the $A^2$ term at zero detuning, $\omega_\text{0} = \omega_\text{a}$, where $\omega_\text{a}$ is the atomic frequency.  The SRPT occurs when there is no $A^2$ term, resulting in a complete frequency softening (a kink) of $\omega_-$ ($\omega_+$) at the phase boundary.

A recent theoretical study has suggested that a magnonic version of the SRPT can occur in ErFeO$_3$ via ultrastrong magnon--spin coupling because the nature of the coupling is an exchange interaction for which there is no $A^2$ term~\cite{Bamba2022}. 
%
Further, terahertz (THz) magnetospectroscopy experiments on a crystal of ErFeO$_3$ have revealed ultrastrong coupling between a magnon mode of ordered Fe$^{3+}$ spins and paramagnetic Er$^{3+}$ spins~\cite{Li2018}.  This system can be modeled by the Dicke Hamiltonian, where the Fe$^{3+}$ magnon mode (the Er$^{3+}$ spins) plays the role of the single cavity mode (the two-level atoms) of the Dicke model; see Fig.\,1A (bottom panel). The $g$ of the Fe$^{3+}$--Er$^{3+}$ coupling exhibited Dicke cooperativity~\cite{Dicke1954,Cong2016}, i.e., $g \propto \sqrt{N}$, where $N$ is the Er$^{3+}$ spin density~\cite{Li2018}. More recently, a short-range atom--atom interaction (Er$^{3+}$--Er$^{3+}$ exchange interaction) has been incorporated for the simulation of an extended Dicke model~\cite{Peraca2023}. However, spectroscopic signatures of the SRPT -- i.e., a polariton frequency softening down to zero and a concomitant change in the other polariton branch -- have not been achieved to date. 

Here we report an unambiguous experimental demonstration of the magnonic SRPT in ErFeO$_3$ through magnetospectroscopy measurements in the THz and gigahertz (GHz) frequency ranges at low temperatures.  We observed that, at the phase boundary between the normal (N) phase and the superradiant (SR) phase, the frequency of a branch of the Er$^{3+}$ electron paramagnetic resonance (EPR) approaches zero while the frequency of a zone-boundary Fe$^{3+}$ magnon displays a kink.  
We developed an extended Dicke model, incorporating the single-ion anisotropy energy of Er$^{3+}$ spins, which accurately reproduces the experimentally observed mode frequencies. The establishment of the magnonic SR phase will enable further experimental explorations of the nonintuitive vacuum-induced ground states predicted for the SR phase~\cite{Makihara2021, Hayashida2023}. 

\subsection*{Expected spectroscopic signatures of the superradiant phase transition}

The standard Dicke model reads,
\begin{equation}
\hat{\mathcal{H}}/\hbar=\omega_\text{0}\hat{a}^\dagger\hat{a}+\omega_\text{a}\left (\hat{S}_z+\frac N2 \right )+\frac{2g}{\sqrt{N}}(\hat{a}^\dagger+\hat{a})\hat{S}_x,
\end{equation}
where $\hat{a}^\dagger$ ($\hat{a}$) is a photon creation (annihilation) operator, $\hat{S}_i$ is a spin operator in the $i$ direction, and $N$ is the number of two-level atoms (Er$^{3+}$ spins).
This model predicts that the SR phase exists at zero temperature when the inequality 
\begin{align}
    g > \frac{\sqrt{\omega_\text{a}\omega_0}}{2}
    \label{eq:SRPTcondition}
\end{align}
is satisfied. In the case of zero detuning ($\omega_\text{a} = \omega_\text{0}$), this condition reduces to $g> \omega_0/2$, i.e., $\eta_\text{c} = 0.5$ is the critical value for the normalized coupling strength $\eta \equiv g/\omega_0$. One can imagine the effect of $g$ is largest when the light and atom  degrees of freedom are on-resonant at zero detuning. Therefore, the standard strategy for realizing the SRPT is to maximize $g$ to reach $\eta_\text{c} = 0.5$ for a fixed $\omega_\text{0}$ while maintaining zero detuning ($\omega_\text{a} = \omega_\text{0}$); see Fig.\,1B. However, even when $\eta < 0.5$, the SRPT can occur if one can reduce $\omega_\text{a}$ to satisfy Eq.\,\ref{eq:SRPTcondition} for fixed $g$ and $\omega_0$. For example, when $\eta = 0.1$ (Fig.\,1C), the inequality in Eq.\,\ref{eq:SRPTcondition} becomes $\nu \equiv \omega_\text{a}/\omega_0 < 0.04$; that is, the SRPT occurs as a function of $\nu$ when it is decreased to the critical value $\nu_\text{c} = 0.04$.  In general, when $\eta < 0.5$ ($\eta > 0.5$), the SRPT occurs $\nu_\text{c} < 1$ ($\nu_\text{c} > 1$), i.e., on the left (right) side of the zero-detuning point when $\omega_\text{a}$ is varied for fixed $g$ and $\omega_0$~\cite{SM}; see Fig.\,S4B.

This nonstandard strategy aptly works for realizing a magnonic SRPT in ErFeO$_3$. The Fe$^{3+}$--Er$^{3+}$ coupling strength $g$ and the Fe$^{3+}$ magnon frequency $\omega_\text{0}$ are nearly independent of the applied magnetic field, $H$, and their ratio is $\eta<1$, while the Er$^{3+}$ EPR frequency $\omega_\text{a}$ strongly depends on the applied magnetic field, via the Zeeman effect. Therefore, applying a magnetic field can tune $\nu$. With realistic values of $g$ and $\omega_\text{0}$ for ErFeO$_3$~\cite{Bamba2022, Peraca2023, Li2022}, the SRPT is expected to occur at a critical magnetic field, $H_\text{c}$, when the temperature, $T$, is sufficiently low ($<$4~K). Notably, the critical temperature, $T_\text{c}$, is maximum when $\omega_\text{a} = 0$, i.e., when $H = 0$. As $H$ increases, $T_\text{c}$ decreases, and hence, the SR phase is transformed into the N phase at $H = H_\text{c}(T)$ when $H$ is varied at a constant temperature; as $T$ is decreased, $H_\text{c}$ monotonically increases from zero to a maximum value at $T=0$, which is a quantum critical point. 
Figure\,1C shows the frequencies of the two polariton branches, $\omega_\pm$, normalized by $\omega_0$, as a function of $\nu$ calculated using the Dicke model in the thermodynamic limit ($N\to\infty$) in the absence of the $A^2$ term; we assumed $\eta = 0.1$, which is a typical value found in the ultrastrong coupling regime~\cite{FornDiaz2019, FriskKockum2019, peraca2020}.


\subsection*{Magnetic structure of ErFeO$_3$}
When $4~\text{K}<T<87~\text{K}$, Fe$^{3+}$ spins are antiferromagnetically ordered along the $c$ axis with a canting toward the $a$ axis by a small angle $\beta$ ($\Gamma_2$ in Bertaut's notation) induced by the Dzyaloshinskii--Moriya (DM) interaction, which produces a weak ferromagnetic moment along the $a$ axis~\cite{Herrmann1963}. As $T$ decreases from 4~K, the N\'eel vector of Fe$^{3+}$ continuously rotates toward the $b$ axis~\cite{Klochan1975}, and paramagnetic Er$^{3+}$ spins develop C-type antiferromagnetic order along the $c$ axis~\cite{Zic2021}. Figure\,2A shows the orthorhombic perovskite structure of ErFeO$_3$ that consists of two Fe$^{3+}$ ($\mathbf{S^{A/B}}$) and two Er$^{3+}$ ($\mathbf{\mathfrak{s}^{A/B}}$) sublattices, described by the space group $D_{2h}^{16}$-$Pbnm$ below 4~K ($\Gamma_{12}$). This phase transition ($\Gamma_2\to\Gamma_{12}$) corresponds to the N\,$\to$\,SR phase transition -- i.e., the appearance of the static coherent electric or magnetic field and atomic polarizations in the context of a photonic SRPT~\cite{Bamba2022}. The two order parameters in the magnonic SRPT can be defined as  $\langle S_y^\text{A/B}\rangle$ and $\langle \mathfrak{s}_z^\text{A}-\mathfrak{s}_z^\text{B}\rangle$~\cite{Peraca2023}. Most importantly, the application of a magnetic field can induce a $\Gamma_{12}\to\Gamma_{2}$ transition~\cite{Kadomtseva1980, Vitebskii1990}, which we utilize in demonstrating the magnonic SRPT.

\subsection*{THz and GHz magnetospectroscopy studies of ErFeO$_3$}
We performed transmission magnetospectroscopy experiments on single crystals of ErFeO$_3$ in the Voigt geometry in the THz and GHz photon frequency ranges to monitor the magnetic field evolution of the upper polariton ($\omega_+$) and lower polariton ($\omega_-$) modes of this Fe$^{3+}$--Er$^{3+}$ hybrid system in Fig.\,1C. The application of a static magnetic field, $H_\text{DC}$, along the $a$ axis continuously tuned the `bare' Er$^{3+}$ EPR frequency $\omega_\text{a}$ via the Zeeman effect, whereas the `bare' Fe$^{3+}$ magnon mode frequency $\omega_\text{0}$ was nearly independent of $H_\text{DC}$. Here, the `bare' frequencies refer to the frequencies of the Fe$^{3+}$ and Er$^{3+}$ modes when they are uncoupled.  The magnetic field moved the system out of the SR phase into the N phase at a critical field of $\mu_0 H_\text{DC} = 1.8$~T at $T = 2$\,K (where $\mu_0$ is the vacuum permeability).

In the THz frequency range (frequencies above 0.25~THz), we used THz time-domain magnetospectroscopy (THz-TDMS)~\cite{Baydin2020} to monitor the $\omega_+$ mode. On the other hand, to monitor the $\omega_-$ mode in the GHz range~\cite{SM}, we used a set of continuous-wave devices (Virginia Diodes, Inc.)\ producing single-frequency microwave radiation at frequencies below 172~GHz. From 33~GHz to 71~GHz, we recorded the intensiy of radiation transmitted through the sample as a function of magnetic field, which exhibited decreases at magnetic resonances. From 74~GHz to 172~GHz, we monitored the sample temperature, which increased at resonance magnetic fields due to resonant absorption of microwave radiation. Through these methods, we were able to locate the resonance frequencies of both the $\omega_+$ and $\omega_-$ modes as a function of magnetic field.

In the THz-TDMS experiments, the magnetic field component of the incident THz wave was set to be parallel to the $a$ axis to access the quasi-antiferromagnetic (qAFM) magnon mode of Fe$^{3+}$~\cite{Li2022} while a static magnetic field was applied along the $a$ axis; see the black box in Fig.\,2B. The THz magnetic field parallel to the net magnetization direction triggered the out-of-phase spin precession in $\mathbf{S^{A/B}}$ through the transient Zeeman torque, as shown in Fig.\,2B. A $b$-cut sample made this configuration possible in the Voigt geometry. In this configuration, we found a pronounced absorption peak at 0.8~THz, which we interpret as the $\omega_+$ mode.
The top panel of Fig.\,2C shows the magnetic field dependence of this mode in an absorption coefficient ($\alpha$) plot~\cite{SM}. The data signals a clear phase transition at 1.8~T as a kink in the field evolution of its frequency. Below 1.8~T, the feature rapidly increased with $H_\text{DC}$, whereas above 1.8~T, it slowly decreased with increasing $H_\text{DC}$. This behavior is consistent with what is shown for the $\omega_+$ mode in Fig.\,1C. 

The bottom three panels of Fig.\,2C display spectra taken in the GHz range, showing a dramatic softening behavior of the $\omega_-$ mode. At 0~T, a broad feature is seen at around 150~GHz, but it rapidly sharpens and red-shifts as the applied magnetic field is increased. 
The frequency of this mode eventually becomes lower than the lower bound of our frequency range but quickly reappears with increasing field, leaving a field gap of 0.2~T between the two peaks at 33~GHz. The middle of these peaks is located at 1.8~T, which agrees with the kink position of the $\omega_+$ mode in the top panel. This softening behavior is consistent with what is shown for the $\omega_-$ mode in Fig.\,1C.  A further increase of the field reveals a Zeeman-type response of the Er$^{3+}$ spins, signifying the absence of antiferromagnetic order in the N phase. The $\omega_-$ mode eventually appears in THz absorption spectra in the top panel above 5.5~T, which confirms the consistency of our two different types of measurements. The less bright line that appears slightly above the $\omega_-$ does not play a significant role in the SRPT~\cite{SM}; see Fig.\,S6.

\subsection*{Phase diagram and polariton frequencies} 
We model our experimental results with a spin Hamiltonian that has been widely studied for rare-earth orthoferrites~\cite{hermann1964magnetic, Yamaguchi2013, Bamba2022}, augmenting it with a term accounting for the anisotropy of the Er$^{3+}$ spins and mapping it to an extended Dicke model; details of the mapping appear in the next section. Figure\,3A displays a mean-field $T$-$H$ phase diagram of the spin Hamiltonian, for $0 \leq T \leq 6$~K and $0 \leq \mu_\text{0} H \leq 3$~T. This phase diagram, consistent with Ref.~\cite{Kadomtseva1980}, shows the SR--phase boundary obtained by using $\langle \mathfrak{s}_{z}^\text{A}-\mathfrak{s}_{z}^\text{B}\rangle$ of Er$^{3+}$ spins as the order parameter. One can also use $\langle S_{y}^\text{A/B}\rangle$ of Fe$^{3+}$ to construct the same phase boundary~\cite{SM}, since the two subsystems simultaneously order (Fig.\,S7). The spin configurations are shown as an inset in Fig.\,3A.

Above 4~K at 0~T, the system hosts antiferromagnetically ordered Fe$^{3+}$ spins while Er$^{3+}$ spins are paramagnetic (Fig.\,2A). As we cool down the system, the $b$ axis component of Fe$^{3+}$ spins becomes finite through the rotation of the N\'eel vector around the $a$ axis. This rotation behavior has been evidenced by nuclear magnetic resonance experiments~\cite{Klochan1975}. At the same time, Er$^{3+}$ spins antiferromagnetically order along the $c$ axis (Fig.\,2A). When we apply an external magnetic field along the $a$ axis stronger than $H_\text{c}$, we break the antiferromagnetic ordering of Er$^{3+}$ spins and restore the Fe$^{3+}$ N\'eel vector along the $c$ axis. These occur simultaneously due to the Fe$^{3+}$--Er$^{3+}$ antisymmetric exchange interaction that couples spins and magnons in the Dicke model as we discuss in the following section.

Figure\,3B shows the frequencies of both polariton branches, calculated based on the spin Hamiltonian, below (2\,K, left panel) and above (10\,K, right panel) the critical temperature as a function of $H$, together with experimental results from Fig.\,2C. At 2~K, good agreement was obtained by fitting Dicke model parameters~\cite{SM}. In particular, our calculations reproduce the observed $\omega_+$ kink and $\omega_-$ softening. At 10~K, we calculated the frequencies with the parameters obtained at 2~K. The two modes at 10\,K do not display any critical behavior and are much less hybridized. Therefore, the two branches are essentially the qAFM mode of Fe$^{3+}$ spins ($\sim\omega_\text{0}$) and the EPR of Er$^{3+}$ spins ($\sim \omega_\text{a}$), respectively. 

Note that the zero-field value of the Er$^{3+}$ resonance is finite even at 10\,K~\cite{Wood1969}. This zero-field splitting is a result of the lifted degeneracy of the Er$^{3+}$ ground state doublets~\cite{Wood1969, Li2018} due to the symmetric Fe$^{3+}$--Er$^{3+}$ exchange interaction that exerts an internal effective magnetic field on Er$^{3+}$ spins. 
Our fitting extracts the bare $\omega_\text{0}$ and $\omega_\text{a}$ appearing in the extended Dicke model~\cite{SM}, and using these values we indicate the value of $\nu = \omega_\text{a}/\omega_0$ in the top $x$ axis in the left panel, finding the critical ratio $\nu_\text{c} \sim 0.11$. 

\subsection*{Mean field theory and extended Dicke model}
The system is described by a spin Hamiltonian on a bipartite lattice with a unit cell consisting of two $\text{Er}^{3+}$ and two $\text{Fe}^{3+}$ spins \cite{Li2018,Bamba2022}.
Nearest-neighboring spins interact via direct magnetic exchange.
Owing to strong spin-orbit coupling in $\text{Fe}^{3+}$, nearest neighbors of $\text{Fe}^{3+}$-spins also have anti-symmetric Dzyaloshinskii--Moriya (DM) interactions between them.
Both the $\text{Fe}^{3+}$ and $\text{Er}^{3+}$ are magnetically anisotropic, preferring the $xz$-plane in our notation.
Although the physical system has four $\text{Fe}^{3+}$ ions in a unit cell, the effective two-sublattice model is able to reproduce the resonance frequencies accurately since the $\text{Fe}^{3+}$-anisotropic energies are much smaller than the anti-symmetric (DM) interaction energy~\cite{Herrmann1963,hermann1964magnetic}. 
The  Hamiltonian describing these effects is
\begin{equation}\label{eq:ham_basic}
    \mathcal{H} = \mathcal{H}_\text{Fe} + \mathcal{H}_\text{Er} + \mathcal{H}_\text{Er-Fe},
\end{equation}
where
\begin{equation}\label{eq:ham_Fe}
    \begin{aligned}
    \mathcal{H}_\text{Fe} = &\sum_{s = \text{A},\text{B}}\sum_{i=1}^{N_0} \mu_\text{B} \mathfrak{g}^{x}_\text{Fe} S^s_{i,x} B_x^\text{DC} + J_\text{Fe}\sum_{\braket{i,i'}}\mathbf{S}_i^\text{A}\cdot \mathbf{S}_{i'}^\text{B}\\
     &-D_\text{Fe}^y \sum_{\braket{i,i'}} \left(S_{i,z}^\text{A} S_{i',x}^\text{B} - S_{i',z}^\text{B} S_{i,x}^\text{A} \right) \\
     &-\sum_{s = \text{A},\text{B}}\sum_{i=1}^{N_0}\left(A^x_\text{Fe} (S_{i,x}^s)^2+A^z_\text{Fe} (S_{i,z}^s)^2\right),
    \end{aligned}
\end{equation}
\begin{equation}\label{eq:ham_Er}
    \begin{aligned}
    \mathcal{H}_\text{Er} = &\sum_{s = \text{A},\text{B}} \sum_{i=1}^{N_0} \mu_\text{B}  \mathfrak{g}_\text{Er}^x \mathfrak{s}_{i,x}^s  B_x^\text{DC} + J_\text{Er} \sum_{\braket{i,i'}} \bm{\mathfrak{s}}_i^\text{A} \cdot \bm{\mathfrak{s}}_{i'}^\text{B} \\
    &-\sum_{s = \text{A},\text{B}}\sum_{i=1}^{N_0}\left(A^x_\text{Er} (\mathfrak{s}_{i,x}^s)^2 + A^z_\text{Er}(\mathfrak{s}_{i,z}^s)^2\right),
    \end{aligned}
\end{equation}
and
\begin{equation}\label{eq:ham_int}
\mathcal{H}_\text{Er-Fe} = \sum_{i=1}^{N_0} \sum_{s,s' = \text{A,B}}\left[J \bm{\mathfrak{s}}_i^s\cdot\mathbf{S}_i^{s'} + \mathbf{D}^{s,s'}\cdot(\bm{\mathfrak{s}}_i^s\times \mathbf{S}_i^{s'})\right].
\end{equation}
The $\mathbf{S}_i^{\text{A}/\text{B}}$ and $\bm{\mathfrak{s}}_i^{\text{A}/\text{B}}$ correspond to $\text{Fe}^{3+}$ ($S=5/2$)  and $\text{Er}^{3+}$ ($\mathfrak{s}=1/2$) spin operators at the $i^\text{th}$-site in the A/B sublattice; $\sum_{\braket{i,i'}}$ represents a sum over the appropriate nearest neighbors;  $N_0$ is the number of unit cells; $\mu_\text{B}$ is the Bohr magneton; $J_\text{Er}$, $J_\text{Fe}$, and $J$ are the direct exchange coupling strengths between Er$^{3+}$--Er$^{3+}$, Fe$^{3+}$--Fe$^{3+}$ and Fe$^{3+}$--Er$^{3+}$ spins, respectively; $D^y_\text{Fe}$ and $\mathbf{D}^{s,s'}$ are the DM-coupling strength between Fe$^{3+}$--Fe$^{3+}$ and Fe$^{3+}$--Er$^{3+}$ spins, respectively; $A^{x/z}_\text{Fe}$ and $A^{x/z}_\text{Er}$ are the anisotropy strengths of $\text{Fe}^{3+}$ and $\text{Er}^{3+}$ spins respectively along the $x/z$ axes; and $\mathfrak{g}_{\text{Fe}/\text{Er}}^x$ are the $x$ component of the Landé g-factor tensors for $\text{Fe}^{3+}$ and $\text{Er}^{3+}$, which are assumed to be diagonal.

We included $\text{Er}^{3+}$-anisotropy terms in Eq.~\ref{eq:ham_Er} which were neglected in previous studies~\cite{Li2018,Bamba2022}. 
However, magnetization measurements~\cite{Bar'yakhtar1977,Dan'shin1986} suggest that $\text{Er}^{3+}$-anisotropy is strong in this material.
The inclusion of the terms is further justified by its large value compared to the exchange interaction between the $\text{Er}^{3+}$-spins, i.e., $A^{x/z}_{\text{Er}}>J_{\text{Er}}$. 
Including the $\text{Er}^{3+}$-anisotropy terms in the Hamiltonian produces a noticeably improved fit to the experimental data (Figs.\,3B and S5). 
This is further elaborated in the Supplementary Material~\cite{SM}.

We calculated the $\omega_{\pm}$ by a combination of mean-field theory and linearization of the Heisenberg equations of motion of the spin operators, as in Ref.~\cite{Bamba2022}. 
Assuming that the average values of the spin operators, $\mathbf{S}^\text{A/B}$ and $\bm{\mathfrak{s}}^\text{A/B}$, are the same in all the unit cells,
 we self-consistently solved for the twelve mean-fields, $\mathbf{\bar{S}}^\text{A/B}$ and $\bm{\bar{\mathfrak{s}}}^\text{A/B}$ in thermal equilibrium.
The resonance frequency were obtained by linearizing the   Heisenberg equations of motion of these twelve spin operators around the mean-field values and solving the resulting linear differential equations \cite{SM}.

The $\text{Er}^{3+}$ parameters in Eq.~\ref{eq:ham_Er}, namely $J_{\text{Er}}$, $\mathfrak{g}_{\text{Er}}^x$, $A_{\text{Er}}^x$, $A_{\text{Er}}^z$,  as well as $A_{\text{Fe}}^z$ and $\mathfrak{g}_{\text{Fe}}^x$ in Eq.~\ref{eq:ham_Fe}, are used as fitting parameters to the spectroscopic data at $T=2\,K$ (Fig.\,3B).
However, the same set of parameters reasonably reproduces the spectroscopic data at 10\,K (Fig.\,S5). 
The rest of the parameters in Eqs.~\ref{eq:ham_Fe} and \ref{eq:ham_int} are the same as those used in Ref.~\cite{Bamba2022}. 
Details of the fitting process and the parameter values are provided in \cite{SM}.

The spin Hamiltonian of $\text{ErFeO}_3$ can be accurately mapped to an extended Dicke Hamiltonian, which provides a natural explanation of the superradiant phase transition. 
Using spin-wave theory, the $\text{Fe}^{3+}$ spins $S^\text{A/B}_i$ are re-written in terms of a set of magnonic creation and annihilation operators ($a_{\text{qFM}}$, $a_{\text{qAFM}}$, $a_{\text{FM}}^\dagger$, $a_{\text{FM}}^\dagger$ ) \cite{Herrmann1963,Bamba2022}, where the "qFM" ("qAFM") corresponds to the $k=0$ ($k=\pi$) quasi-momentum modes.
Following Ref.~\cite{Bamba2022}, we also make a collective spin approximation for the $\text{Er}^{3+}$-spins,
\begin{align}
    \mathfrak{s}^{\text{A}/\text{B}}_{i,\xi} &= \frac{1}{N_0}\sum_{i=1}^{N_0} \mathfrak{s}_{i,\xi}^{\text{A}/\text{B}}  \\
    &\equiv \frac{1}{N_0}\Sigma_{\xi}^{\text{A}/\text{B}},
\end{align}
where $\xi=x,y,z$ and the last line defines two sets of spin operators $\Sigma_\xi^{\pm} = \sum_i \left(\mathfrak{s}_{\xi,i}^\text{A} \pm \mathfrak{s}_{\xi,i}^\text{B}\right)$.
The magnon representation of $\text{Fe}^{3+}$ is justified owing to the strong exchange interaction $J_{\text{Fe}} = 4.96$\,meV, making it highly dispersive, while the  $\text{Er}^{3+}$ exchange interaction, $J_{\text{Er}}=0.0132$\,meV, is more than two orders magnitude smaller. Hence the $\text{Er}^{3+}$-spins are  well-described by the collective spin approximation, and they couple only to the qAFM and qFM iron modes.
The explicit form of the extended Dicke Hamiltonian is
\begin{equation}\label{eq:Dicke}
\begin{aligned}
    \mathcal{H}_\text{Dicke} = & \sum_{\text{m}:\{\text{qFM},\text{qAFM}\}}\hbar\omega_{\text{m}} a_{\text{m}}^\dagger a_{\text{m}} +  E_x \Sigma_x^+ + E_y \Sigma_y^+ + \mu_\text{B}\mu_0\mathfrak{g}_\text{Er}^x H_x^\text{DC}\Sigma_x^+ + \frac{z_\text{Er}J_\text{Er}}{N_0}\bm{\Sigma}^\text{A}\cdot\bm{\Sigma}^\text{B} \\
     &-\sum_{\xi=x,z}\sum_{s=A,B}\frac{A^\xi_\text{Er}}{N_0}(\Sigma_\xi^\text{s})^2  + \frac{g_x}{\sqrt{N_0}} \left(a_{\text{qAFM}}^\dagger + a_{\text{qAFM}}\right)\Sigma_x^+  + \frac{ig_y}{\sqrt{N_0}}\left(a_{\text{qFM}}^\dagger - a_{\text{qFM}}\right)\Sigma_y^+\\
     & + \frac{g_{y'}}{\sqrt{N_0}}\left(a_{\text{qAFM}}^\dagger + a_{\text{qAFM}}\right)\Sigma_y^- + \frac{ig_z}{\sqrt{N_0}}\left(a_{\text{qAFM}}^\dagger - a_{\text{qAFM}}\right)\Sigma_z^- + \frac{g_{z'}}{\sqrt{N_0}}\left(a_{\text{qFM}}^\dagger + a_{\text{qFM}}\right)\Sigma_z^+.
    \end{aligned}
\end{equation}
Since the  $\text{Er}^{3+}$ decouple from the magnon modes for $k\neq0,\pi$, we omit these in  Eq.~\ref{eq:Dicke}.
The Supplementary Materials shows the derivation of Eq.~\ref{eq:Dicke} from Eqs.~\cref{eq:ham_basic,eq:ham_Fe,eq:ham_Er,eq:ham_int} and the equations for the Dicke model parameters ($\omega_\text{qFM}$, $\omega_{\text{qAFM}}$,  $E_x$, $E_y$, $g_x$, $g_y$, $g_{z}$, $g_{y'}$, and $g_{z'}$) \cite{SM}.

Similar to Ref.~\cite{Bamba2022}, the temperature dependence is incorporated through the dilution factor  
\begin{equation}
    x = \tanh\left[\frac{|E_x + \mu_\text{B}\mu_0\mathfrak{g}^x_{\text{Er}}H_x^{\text{DC}}|}{2k_\text{B} T}\right],
\end{equation}
where $k_\text{B}$ is the Boltzmann constant.
The only difference between Eq.~\ref{eq:Dicke} and the extended Dicke model of Ref.~\cite{Bamba2022} is the $\text{Er}^{3+}$-anisotropy terms.


\subsection*{Discussion} 
We provided spectroscopic evidence that the $\Gamma_{12}\to\Gamma_{2}$ phase transition in ErFeO$_3$ can be understood as the SR$\to$N transition in central to quantum optics~\cite{Bamba2022}. This phase transition can also be regarded as the magnetic transition of the Jahn-Teller type~\cite{Zvezdin1976,Porras2012}. What makes our mode softening distinct from others found in various solids, is the concomitant kink and applicability of the Dicke model. While our softening resembles a softening at a spin-flop transition, as found in MnF$_2$ (easy-axis antiferromagnet)~\cite{Hagiwwara1999}, the softening occurs without any concomitant kink since only one magnetic sub-lattice exists in the material. Furthermore, this softening occurs only when the external magnetic field is applied parallel to the N\`eel vector (easy-axis), in contrast to our work where the field is applied perpendicular to the N\`eel vector (hard-axis). In conclusion, our softening does not stem from a simple spin-flop transition. 

Our work is readily applied to other solids, such as other rare-earth orthoferrite or orthochromite compounds, where two different magnetic sub-lattices strongly interact with each other, which host different types of phase transitions~\cite{Belov1976,balbashov1995submillimeter}. One can simulate different types of Dicke models or explore novel quantum vacuum phenomena at or in the SR phase by judiciously choosing candidate materials. We suggest two necessary conditions that must be satisfied for this analogy to be valid as a general guidance. First, one should find evidence that $g$ exhibits the Dicke cooperative enhancement ($g\propto\sqrt{N}$). Second, at the superradiant phase boundary, one should find a simultaneous change in two magnetic sub-lattices that can be revealed by a kink and softening in spectroscopic measurements. Theoretical mapping of a spin Hamiltonian into Dicke models should be possible. 


\subsection*{Conclusion}
We observed the spectroscopic signatures of the magnonic SRPT in ErFeO$_3$ in thermal equilibrium. Our work demonstrates the long-sought SRPT predicted by Hepp and Lieb in the Dicke model without the $A^2$ term~\cite{HEPP1973360} The magnon--spin system opens up the possibilities to explore novel quantum vacuum phenomena predicted in the superradiant phase. At the superradiant phase boundary of the Dicke model, a two-mode ground-state becomes perfectly squeezed~\cite{Hayashida2023}. This suggests that ErFeO$_3$ in a magnetic field offers a unique opportunity to achieve large-scale quantum entanglement essential for quantum information science, intrinsically robust against external noise -- equilibration with the thermal bath provides a passive ``error correction'' of decoherence events. Furthermore, our work provides insights into achieving photonic SRPTs through an exchange pathway, as well as a novel way to discover and control condensed matter phases based on powerful concepts developed in quantum electrodynamics.


\bibliography{scibib}

\bibliographystyle{Science}

\section*{Acknowledgments}
D.K.\ and J.K.\ acknowledge support from the U.S.\ Army Research Office (through Award No.\ W911NF2110157), the Gordon and Betty Moore Foundation (through Grant No.\ 11520), and the Robert A.\ Welch Foundation (through Grant No.\ C-1509).  K.H.\ and J.K.\ acknowledge support from the W.\ M.\ Keck Foundation (through Award No.\ 995764). D.K. and J.K. acknowledge support from Global Institute for Materials Research Tohoku and International Collaborations with Institute for Materials Research, Tohoku University. K.H. acknowledges support from  the National Science Foundation (PHY-1848304). M.B.\ acknowledges support from the Japan Society for the Promotion of Science (through Grant No.\ JPJSJRP20221202). L.L., J.-M.P., D.C., R.K. and J.W. were supported by the US Department of Energy, Office of Basic Energy Science, Division of Materials Sciences and Engineering (Ames National Laboratory is operated for the US Department of Energy by Iowa State University under contract no. DE-AC02-07CH11358).

\section*{Author contributions}
J.K. conceived and supervised the project. D.K. conceived detailed experimental plans, performed THz measurements and GHz temperature measurements, analyzed all experimental data, and prepared the manuscript under the supervision and guidance of J.K. S.D. and H.-T.W. wrote the theory section in the manuscript, introduced the anisotropy term, and performed fitting under the supervision and guidance of M.B. and K.H. M.B. developed the theoretical model. X.M. and W.Y. grew, cut, and characterized the crystals used in the experiments under the guidance of S.C. J.D. polished the samples. J.-M.P. performed THz measurements and GHz temperature measurements. L.L., J.-M.P, D.C. and R.K. built the THz setup under guidance of J.W. H.N., S.K. and D.K. performed GHz transmission measurements. All authors discussed the results and commented on the manuscript.

\section*{Competing interests}
None declared.

\section*{Data and materials availability}
All data are available in the manuscript or the supplementary materials.

\section*{Supplementary materials}
Materials and Methods\\
Supplementary Text\\
Figs. S1 to S7\\
Table S1\\
References \textit{(1-9)}


\clearpage
\begin{figure}[t]
		\centering
		\includegraphics[width=1\linewidth]{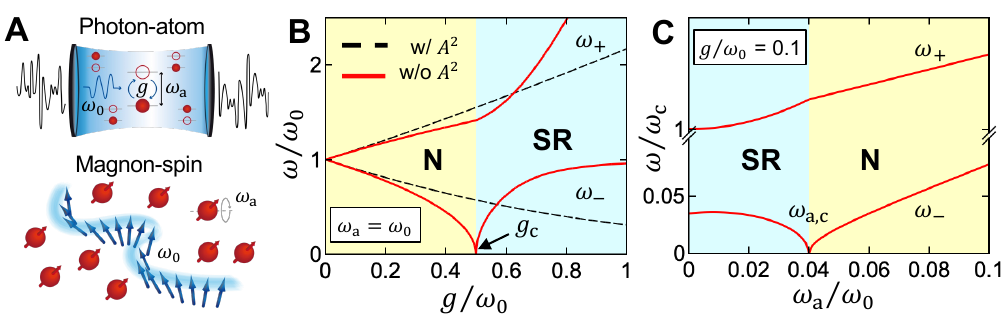}
	\end{figure}
\noindent \textbf{Fig.\,1. Comparison between a light--matter system and a magnon--spin system for the Dicke superradiant phase transition (SRPT).} (\textbf{A}) (Top panel)~A light--matter hybrid system with coupling strength $g$ realized in a single-mode cavity with frequency $\omega_\text{0}$ containing an ensemble of two-level atoms with transition frequency $\omega_\text{a}$. (Bottom panel)~A magnon--spin hybrid system realized in ErFeO$_3$. A single-mode magnon excitation of Fe$^{3+}$ (an ensemble of Er$^{3+}$ spins) plays the role of single-mode cavity photons (two-level atoms) in the Dicke model. (\textbf{B})~Normalized frequencies of the upper-polariton ($\omega_{+}$) and lower-polariton ($\omega_{-}$) modes as a function of $g/\omega_\text{0}$ calculated using the Dicke model without (red solid lines) and with (black dashed lines) the $A^2$ term  at zero-detuning ($\omega_\text{0} = \omega_\text{a}$). When the lower-polariton frequency reaches zero, an SRPT occurs between the normal (N) and superradiant (SR) phases. With the $A^2$ term, the lower-polariton frequency asymptotically approaches zero in the $g/\omega_\text{0} \rightarrow \infty$ limit, thereby forbidding the SRPT.  (\textbf{C})~Normalized frequencies of the upper-polariton ($\omega_{+}$) and lower-polariton ($\omega_{-}$) modes as a function of $\omega_\text{a}/\omega_\text{0}$ calculated using the Dicke model without the $A^2$ term with $g/\omega_\text{0} = 0.1$. When the lower-polariton frequency reaches zero, the system crosses the phase boundary. 

\clearpage
\begin{figure}[t]
		\centering
		\includegraphics[width=1\linewidth]{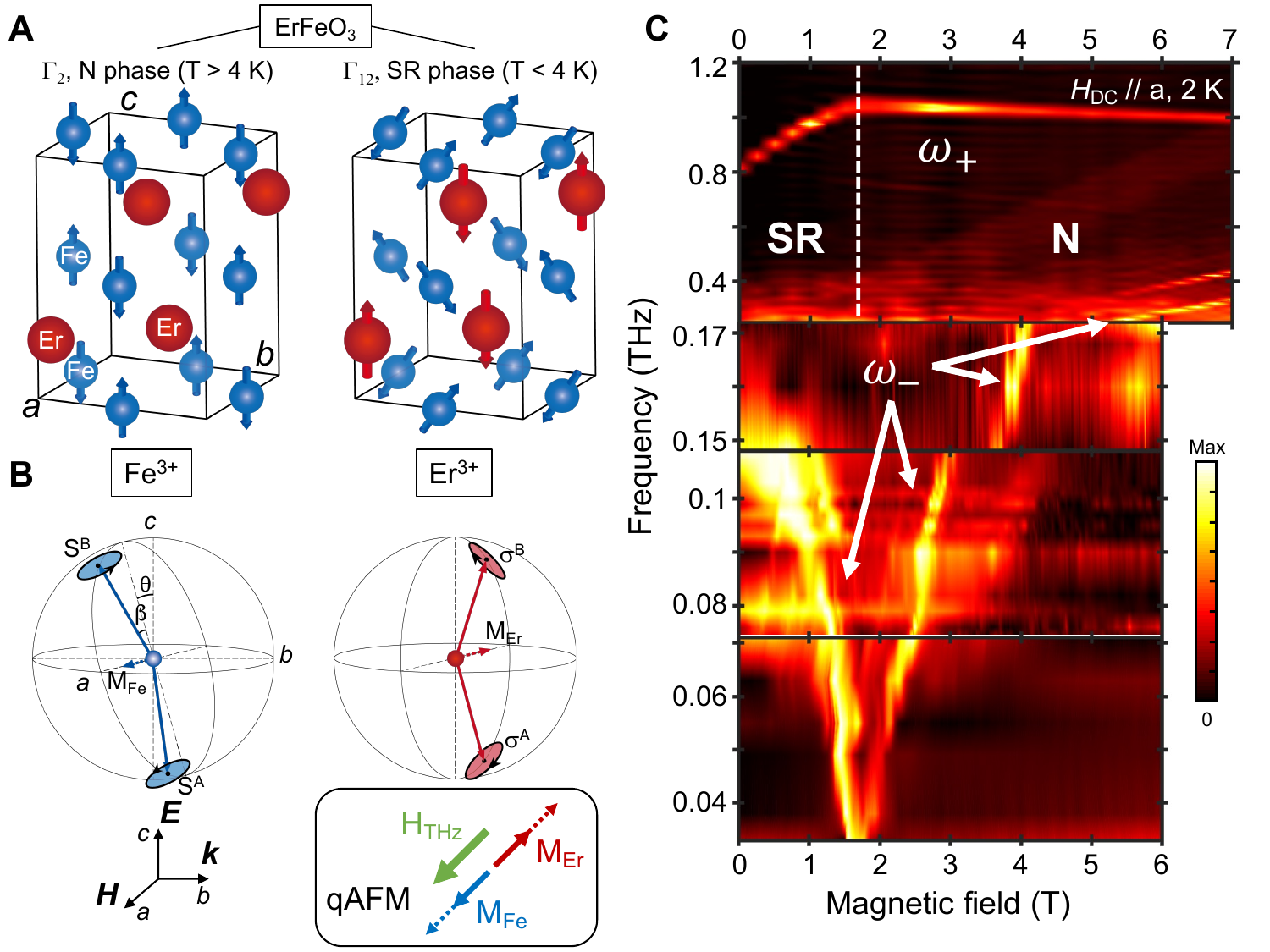}
	\end{figure}
 
\noindent {\textbf{Fig. 2. Spectroscopic evidence for the magnonic SRPT in ErFeO$_3$.}} (\textbf{A})~A schematic of the magnetic structure of ErFeO$_3$ in the $\Gamma_{12}$ or superradiant phase ($T < T_{\text{N}}^{\text{Er}}$). Er$^{3+}$ spins (blue vectors) become antiferromagnetically aligned at low temperatures, while the N\'eel vector of the Fe$^{3+}$ spins rotates toward the $b$ axis. (\textbf{B})~Spin dynamics of the qAFM of Fe$^{3+}$ spins and Er$^{3+}$ spins in the $\Gamma_{12}$ phase triggered by a THz magnetic field polarized along the $a$ axis. The magnitude of the net magnetization (M$_\mathrm{Fe}$ and M$_\mathrm{Er}$) oscillate (black box). The Fe$^{3+}$ spins are ordered along the $c$ axis and cant toward the $a$ axis ($\beta$~=~8.5~mrad). The plane where the Fe$^{3+}$ spins lie is $\theta$~=~49\textdegree~off from the $ac$ plane. (\textbf{C}) (Top panel)~Absorption coefficient spectra as a function of magnetic field in THz-TDS, showing a kink in the upper-polariton. Here the maximum (minimum) value is 60~cm$^{-1}$ (0~cm$^{-1}$). (Two middle panels)~Temperature spectra in the sample upon CW GHz illumination with a base temperature of 2~K. The maximum temperature change is less than 90\,mK. Here all spectra are scaled from 0 to 1. (Bottom panel)~Negative transmission spectra. Here all spectra are scaled from 0 to 1. The GHz spectra in the bottom three panels show a softening in the lower-polariton frequency, which together with the kink in the upper-polariton mode, demonstrates the magnonic SRPT; see Fig.\,1C. 

\clearpage
\begin{figure}[t]
\centering
\includegraphics[width=1\linewidth]{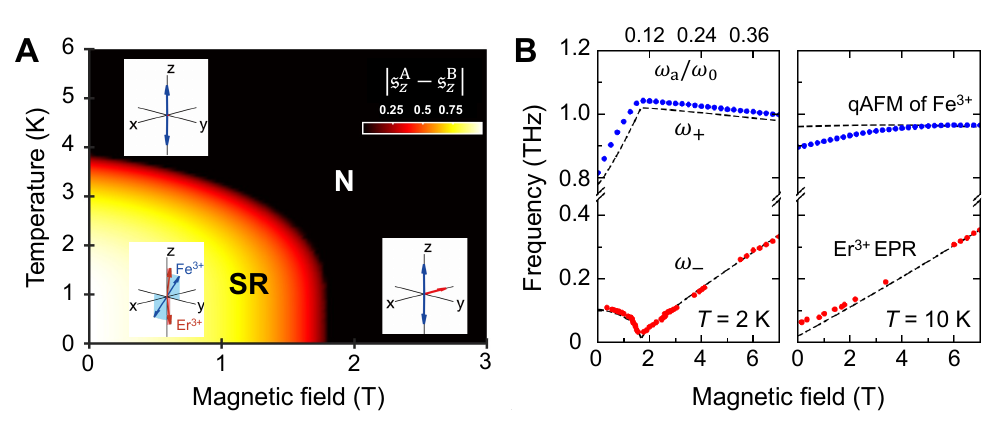}
\label{fig:exp_theory} 
\end{figure}

 \noindent {\textbf{Fig. 3. Mean-field calculation for the spin Hamiltonian of ErFeO$_3$ in $H \parallel a$.} (\textbf{A})~Theoretical $T$-$H$ phase diagram. (Insets) Schematics of the spin configuration in each phase. (\textbf{B}) Resonance frequencies of each spin subsystem as a function of the external magnetic field at (left panel) 2~K and (right panel) 10~K. Dots: experimental results. Dashed lines: calculated resonance frequencies. The error bars show the full-width half-maximum of the peaks. The top $x$ axis on the left panel shows the SRPT occurs at the ratio of $\omega_\text{a}/\omega_\text{0} = 0.11$.}

\appendix



\topmargin 0.0cm
\oddsidemargin 0.2cm
\textwidth 16cm 
\textheight 21cm
\footskip 1.0cm


\newpage
\section*{Supplementary Materials for ``Observation of the Magnonic Dicke Superradiant Phase Transition''} 


\author
{Dasom Kim$^{1,2,3\dagger}$, Sohail Dasgupta$^{4,\dagger}$, Xiaoxuan Ma$^{5,\dagger}$, Joong-Mok Park$^{3}$,\\ Hao-Tian Wei$^{4}$, Liang Luo$^{3}$, Jacques Doumani$^{1,2}$, Xinwei Li$^{6}$, Wanting Yang$^{5}$,\\ Di Cheng$^{3,7}$, Richard H. J. Kim$^{3}$, Henry O. Everitt$^{2,8,9}$, Shojiro Kimura$^{10}$,\\ Hiroyuki Nojiri$^{10}$,  Jigang Wang$^{3,7}$, Shixun Cao$^{5,\ast}$, Motoaki Bamba$^{11}$,\\ Kaden R. A. Hazzard$^{4,8,12}$,  Junichiro Kono$^{2,4,8,13,\ast}$\\
\\
\normalsize{$^{1}$Applied Physics Graduate Program, Smalley--Curl Institute, Rice University,}\\
\normalsize{Houston, TX 77005, USA}\\
\normalsize{$^{2}$Department of Electrical and Computer Engineering, Rice University,}\\
\normalsize{Houston, TX 77005, USA}\\
\normalsize{${^3}$Ames National Laboratory, Ames, IA 50011, USA}\\
\normalsize{$^{4}$Department of Physics and Astronomy, Rice University, Houston, TX 77005, USA}\\
\normalsize{$^{5}$Department of Physics, International Center of Quantum and Molecular Structures,}\\
\normalsize{and Materials Genome Institute, Shanghai University, Shanghai, 200444, China}\\
\normalsize{$^{6}$Department of Physics, National University of Singapore, 117551, Singapore}\\
\normalsize{$^{7}$Department of Physics and Astronomy, Iowa State University, Ames, IA 50011, USA}\\
\normalsize{$^{8}$Smalley–Curl Institute, Rice University, Houston, TX 77005, USA}\\
\normalsize{$^{9}$DEVCOM Army Research Laboratory-South, Houston, TX 77005, USA}\\
\normalsize{$^{10}$Institute for Materials Research, Tohoku University, Sendai 980-8577, Japan}\\
\normalsize{$^{11}$Department of Physics, Yokohama National University, Yokohama 240-8501, Japan}\\
\normalsize{$^{12}$Department of Physics and Astronomy, University of California, Davis, CA 95616, USA}\\
\normalsize{$^{13}$Department of Materials Science and NanoEngineering, Rice University,}\\
\normalsize{Houston, TX 77005, USA}\\
\\
\normalsize{$^\ast$To whom correspondence should be addressed; E-mail: sxcao@shu.edu.cn, kono@rice.edu}\\
\normalsize{$^\dagger$These authors contributed equally.}
}


\date{}


\baselineskip24pt

\maketitle 

\newpage

\section*{Materials and Methods}

\subsection*{\underline{Sample preparation}}
Polycrystalline ErFeO$_3$ was first synthesized by a conventional solid-state reaction method using Er$_2$O$_3$ (99.9\%) and Fe$_2$O$_3$ (99.98\%) powders. According to the stoichiometric ratio, the original reagents were weighed carefully and pulverized with moderate anhydrous ethanol in an agate mortar. Mixtures were sintered at 1300 \textdegree C for 1000 minutes and then cooled to room temperature. The sintered powders were thoroughly reground and pressed into a rod that is 70~mm in length and 5~-~6~mm in diameter by a Hydrostatic Press System (Riken Seiki CO. Ltd, model HP-M-SD-200) at 70~MPa, and then sintered again at 1300~°C for sufficient reaction. Single crystal samples were then grown by an optical floating zone furnace (FZT-10000-H-VI-P-SH, Crystal Systems Corp; heat source: four 1~kW halogen lamps). Conditions like the melting power and the rate of sample rotation were stabilized and controlled in the molten zone.

\subsection*{\underline{Gigahertz magnetospectroscopy}}

\subsubsection*{Thermal detection of electron paramagnetic resonances in static magnetic fields under GHz radiation}
We performed GHz magnetospectroscopy measurements in the Voigt geometry as we described in the main text. Figure S1A shows our setup used for thermal detection of magnetic resonances of Er$^{3+}$ spins. We generate GHz continuous waves (CW) using a diode (70~-~110~GHz, 25~mW, Virginia diode, WR10SGX). For the frequency range (140~-~220~GHz), a frequency multiplier (3.2~mW, WR5.1x2) was added. The GHz radiation was collimated by a 90\textdegree~off-axis parabolic mirror and shined on the sample to ensure that the whole area of the sample (4~mm in diameter) was uniformly illuminated. A wire-grid polarizer was used to define the GHz magnetic field polarization. The $c$-cut sample (800-$\mu$m-thick) was placed in a dry magneto-optical cryostat (Quantum Design, OptiCool) at 2~K and static magnetic fields $\mu_0 H \parallel a$ up to 6 T. A temperature sensor was located about 5~mm far from the sample and they sit on the same Cu plate. For each fixed frequency, we swept the magnetic fields to 6~T with the step size of 0.02~T. At each magnetic field, we waited for 5~s for the temperatures to be stabilized. We selected frequencies that show a temperature change of $30~\text{mK}<\Delta T<90~\text{mK}$ at the peak temperatures.

Figure S1B shows an example of the thermal detection. With 90~GHz illumination, as we increase the magnetic field, the sample temperature increases at some magnetic fields. This is because when the incident photon energy coincides with the transition energy of the spin resonances, non-radiative electron relaxation emitting phonons can happen, thereby increasing the lattice temperature. This technique has been used for observing cyclotron resonances in the microwave regime~\cite{Zhang2018}. One of the advantages of using this method over the transmission measurement is that one can avoid the notorious standing wave effects that make the transmission curve more complicated. 

Figure S2A shows the measured curves at 2~K through this method with frequencies ranging from 74 to 172~GHz which are 0-to-1 scaled. These data are used to generate the middle two panels in Fig.\,2C. Above 95\,GHz, side peaks appear around 1.5\,T. They are hybridized modes of the Er$^{3+}$ mode and a Fabry-P\'erot cavity mode defined by the crystal itself~\cite{Kritzell2023}. For 10~K measurements, this method does not work since the cryostat can maintain the temperature at 10~K even with GHz radiation. At 2~K, however, the cryostat is using its maximum cooling capability, and thereby any additional thermal load results in increases in the temperature.

\subsubsection*{Transmission in a pulsed magnetic field}
High-magnetic field electron paramagnetic resonance (EPR) measurements in pulsed magnetic fields with fixed frequencies ranging from 33 to 71~GHz and 63 to 190~GHz were performed in the temperatures at 2 and 10~K, respectively. Pulsed magnetic fields up to 5~T were applied. Magnetic fields were applied parallel to the $a$ axis. All curves have been 0 to 1 re-scaled.

At 2~K, we were able to observe the resonances in transmission spectra since the sample used for the pulsed magnetic field experiment was thinner (200 $\mu$m) than the incident wavelengths, allowing us to avoid the standing waves effect inside the sample; Fig.\,S2A. Note that the dielectric constant of ErFeO$_3$ is about 30. These data are used to generate the bottom panel in Fig.\,2C. Figure S2B shows transmission spectra at 10~K. These data are used in Fig.\,3B (right panel). Here, since the thermal energy of 10~K (208~GHz) is higher than the transition energies, the population in the upper level is high and thereby the transmission changes are minute. This results in the high background noise at low frequencies after we do the 0-to-1 normalization.

\subsection*{\underline{Terahertz magnetospectroscopy}}
\subsubsection*{Setup description}
We performed THz time-domain magnetospectroscopy (THz-TDMS) measurements in the Voigt geometry as we described in the main text. The $b$-cut sample (1162-$\mu$m-thick) is placed in a dry magneto-optical cryostat (Quantum Design, OptiCool) with variable temperatures $T$ between 2 and 300~K and static magnetic fields $\mu_0 H$ up to 7~T. We generate THz pulses via optical rectification using a Ti:sapphire amplifier (795~nm, 0.2~mJ, 40~fs, 1~kHz, Spectra-Physics, Spitfire Ace Pro XP) as a laser source that pumps a 1-mm-thick (110) zinc telluride (ZnTe) crystal, while detection is accomplished through electro-optical sampling in another ZnTe crystal with the same thickness.

\subsubsection*{Index of Refraction and Absorption Coefficient}
We extracted the absorption coefficient spectra using complex indices of refraction of a sample obtained by THz-TMDS~\cite{Peraca2023}. We define $\tilde{E}_0(\omega)$ as the Fourier transform of an incoming THz pulse $E_0(t)$ illuminated on a homogeneous dielectric slab of thickness $d$ (the sample thickness). We neglected the ghost reflections from the backside of the sample (the Fabry-P\'erot effect) since they are well separated in time-domain from the main transmitted THz pulse and excluded in our analysis. We define the THz electric field transmitting with and without the sample as the sample electric field $\tilde{E}_\text{s}(\omega)$ and reference electric field $\tilde{E}_\text{r}(\omega)$, respectively. Each transmitted electric field can be written as \cite{Naftaly2015,Duvillaret1996}:
	\begin{align}
		\tilde{E}_{\text{r}}(\omega) = \tilde{t}_{13}(\omega)\tilde{P}_{\text{air}}(\omega,d)\tilde{E}_0(\omega) \label{Er} \\
		\tilde{E}_{\text{s}}(\omega) = \tilde{t}_{12}(\omega)\tilde{P}_{\text{s}}(\omega,d)\tilde{t}_{23}\tilde{E}_0(\omega) \label{Es}
	\end{align}
where $\tilde{t}_{jk} = \frac{2\tilde{n}_j}{\tilde{n}_j+\tilde{n}_k}$ is the complex Fresnel transmission coefficient between mediums $j$ and $k$, $\tilde{P}_j(\omega,d_j) = e^{ik_0d_j\tilde{n}_j} = e^{i(\omega d_j/c)\tilde{n}_j}$ is the propagator through medium $j$, and the subscripts $\text{air}$, $\text{r}$, and $\text{s}$ refer to air, reference, and sample, respectively. The ratio between $\tilde{E}_{\text{r}}(\omega)$ and $\tilde{E}_{\text{s}}(\omega)$ is the transfer function $\tilde{H}(\omega)$, and it follows from Eqs.~\ref{Er}~and~\ref{Es} that:
	\begin{align}
		\tilde{H}(\omega) = \frac{\tilde{E}_{\text{s}}(\omega)}{\tilde{E}_{\text{r}}(\omega)} = \frac{\tilde{t}_{12}\tilde{t}_{23}}{\tilde{t}_{13}}\frac{\tilde{P}_{\text{s}}(\omega,d)}{\tilde{P}_{\text{air}}(\omega,d)} = \frac{2\tilde{n}_2(\tilde{n}_1+\tilde{n}_3)}{(\tilde{n}_1 + \tilde{n}_2)(\tilde{n}_2+\tilde{n}_3)} e^{i(\omega d/c)(\tilde{n}_\text{s}-1)}
		\label{H}
	\end{align}
Since we are dealing with a bulk sample that does not have any substrate, the surrounding mediums can be taken as air by setting $\tilde{n}_1 = \tilde{n}_3 = 1$ in Eq.\,\ref{H}. With this simplification, the pre-factor before the exponential term becomes $\frac{4\tilde{n}_\text{s}}{(\tilde{n}_\text{s}+1)^2}$ with $\tilde{n}_2 = \tilde{n}_\text{s}$. Furthermore, we set $\tilde{n}_\text{s} = n_\text{s}(\omega)$ for $\tilde{t}_{jk}$ and solve Eq.\,\ref{H} for $\tilde{n}_\text{s} = n_\text{s}(\omega) + i\kappa_\text{s}(\omega)$ in the exponential term. Here, $n_\text{s}(\omega)$ is the index of refraction of the sample, and $\kappa_\text{s}(\omega)$ its extinction coefficient. This approximation is based on the fact that the absorption at the sample surface is negligible compared to the exponential term. We obtain:
	\begin{align}
		\label{H2}
		\tilde{H}(\omega) = \frac{4n_\text{s}(\omega)}{(n_\text{s}(\omega)+1)^2}e^{i(\omega d/c)(\tilde{n}_\text{s}-1)} =  \frac{4n_\text{s}(\omega)}{(n_\text{s}(\omega)+1)^2}e^{i(\omega d/c)(n_\text{s}(\omega) -1)}e^{-(\omega d/c)\kappa_\text{s}(\omega)}
	\end{align}
The magnitude and argument of Eq.\,\ref{H2} read:
	\begin{align}
		\text{arg}[\tilde{H}(\omega)] & = \left(\frac{\omega d}{c}\right)(n_\text{s}(\omega)-1) \rightarrow n_\text{s}(\omega) = 1 + \frac{c}{\omega d}\text{arg}[\tilde{H}(\omega)] \label{nw} \\[5pt]
		|\tilde{H}(\omega)| & = \frac{4n_\text{s}(\omega)}{(n_\text{s}(\omega)+1)^2}e^{-(\omega d/c)\kappa_\text{s}(\omega)} \rightarrow \kappa_\text{s}(\omega) =  -\frac{c}{\omega d}\ln\left[\frac{(n_\text{s}(\omega)+1)^2}{4n_\text{s}(\omega)}|\tilde{H}(\omega)|\right]
	\end{align}
Then the absorption coefficient  $\alpha(\omega)$ reads as follow:
	\begin{align}
		\alpha(\omega) & =  \frac{2\omega}{c}\kappa(\omega) = -\frac{2}{d}\ln\left[\frac{(n_\text{s}(\omega)+1)^2}{4n_\text{s}(\omega)}|\tilde{H}(\omega)|\right] \label{aw}
	\end{align}
To summarize, from $\tilde{E}_{\text{r}}(t)$ and $\tilde{E}_{\text{s}}(t)$, obtained in this work, the transfer function $\tilde{H}(\omega)$ can be calculated as $\tilde{E}_{\text{s}}(\omega)/\tilde{E}_{\text{r}}(\omega)$, and $n(\omega)$ and $\alpha(\omega)$ follow from Eqs.~\ref{nw}~and~\ref{aw}, respectively.

\newpage
\section*{Supplementary Text}

\subsection*{\underline{THz absorption spectra at high temperatures}}
Figure\,S3A shows temperature-dependent absorption spectra of qAFM of Fe$^{3+}$ spins. The kink occurs at 4~K which is the superradiant phase boundary at 0~T. Below this temperature, the Fe$^{3+}$ order parameter $\langle S^\text{A/B}_y\rangle$ becomes finite. Figure\,S3B shows magnetic field dependence of qAFM of Fe$^{3+}$ spins. At 10\,K, only a slight change was observed at low magnetic fields without any signature of the phase transition, consistent with our phase diagram. Meanwhile, two modes that emerge at high fields are Er$^{3+}$ EPR modes. As described in the main text, ErFeO$_3$ can be modeled by the two-sublattice model. This implies we should expect four modes in total: quasi-ferromagnetic (qFM) mode and qAFM for Fe$^{3+}$ spins, and in-phase and out-of-phase EPR modes for Er$^{3+}$ spins. Here, we are considering the relative phase of precession of two Er$^{3+}$ spins.  A detailed derivation follows in the next section and can be found in Ref.~\cite{Bamba2022}. Our theory finds the lowest mode is the out-of-phase mode that is coupled to qAFM, establishing a magnon-spin system. Due to the polarization selection rule described in Fig.\,2B, the qFM mode does not appear in this plot.

\subsection*{\underline{Superradiant phase transition at finite detuning}}
An anti-crossing of two polaritons occurs at the zero-detuning point ($\omega_\text{0} = \omega_\text{a}$). When the normalized coupling strength ($\eta \equiv g/\omega_\text{0}$) reaches the critical value of 0.5, the system undergoes the SRPT, and $\omega_-$ becomes zero. As shown in Fig.\,S4A, for $\eta = 0.5$ the SRPT occurs when $\omega_\text{0} = \omega_\text{a}$. For $\eta = 0.1$ (Fig.\,S4B), a situation  more comparable to ErFeO$_3$,  we find the phase boundary moves to the $\omega_a<\omega_0$, while the anti-crossing still occurs at the zero-detuning point. Thus one can achieve the SRPTs with a small  $\eta$ as long as $\nu \equiv \omega_\text{a}/\omega_0$ is small enough to satisfy the inequality (1) in the main text. By contrast, when $\nu > 1$, the $\eta_\text{c}$ becomes higher than 0.5.


\subsection*{\underline{Resonance frequencies from the spin model}}
We calculate the resonance frequencies from the spin Hamiltonian with the method of Ref.~\cite{Bamba2022}.
The Heisenberg equations of motion of the twelve spin operators are
\begin{align}
    \hbar   \frac{d\bm{\mathfrak{s}}^s}{dt} &= -\bm{\mathfrak{s}}^s \times \mathfrak{g}_{\text{Er}}\mu_\text{B} \mathbf{B}_{\text{Er}}^{s} (\bm{\mathfrak{s}}^\text{A/B},\mathbf{S}^\text{A/B}), \\
    \hbar \frac{d\mathbf{S}^s}{d t} &= -\mathbf{S}^s \times \mathfrak{g}_{\text{Fe}}\mu_\text{B} \mathbf{B}_{\text{Fe}}^s (\bm{\mathfrak{s}}^\text{A/B},\mathbf{S}^\text{A/B}),
\end{align}
dropping the unit cell index due to the assumption that spins are spatially uniform, with $s\in\{A,B\}$, and $\mathbf{B}_{\text{Er}}^{s}$ ($\mathbf{B}_{\text{Fe}}^{s}$)  the mean-fields for the $\text{Er}^{3+}$ ($\text{Fe}^{3+}$) spins,
\begin{align}
   \mathbf{B}_{\text{Er}}^{s} &=  \mathbf{B}^{\text{DC}} + \frac{2 z_{\text{Er}}J_{\text{Er}}}{\mu_\text{B} \mathfrak{g}_{\text{Er}}}\bm{\mathfrak{s}}^{\bar{s}} + \frac{2}{\mu_\text{B}\mathfrak{g}_{\text{Er}}}\sum_{s'=\text{A,B}} \left[  J \mathbf{S}^s - (\mathbf{D}^{s,s'}\times\mathbf{S}^s) -\mathbf{A}_{\text{Er}}\cdot\bm{\mathfrak{s}}^{s}\right],\label{eq:mean-field-er} \\
   \mathbf{B}_{\text{Fe}}^{s} &=  \mathbf{B}^{\text{DC}} + \frac{2 z_{\text{Fe}}J_{\text{Fe}}}{\mu_\text{B} \mathfrak{g}_{\text{Fe}}}\mathbf{S}^{\bar{s}} - \frac{z_{\text{Fe}}}{\mu_\text{B}\mathfrak{g}_\text{Fe}}\mathbf{D}_{\text{Fe}}\times\mathbf{S}^{\bar{s}}+ \frac{2}{\mu_\text{B}\mathfrak{g}_{\text{Fe}}}\sum_{s'=\text{A,B}} \left[  J \mathbf{S}^s - (\mathbf{D}^{s,s'}\times\mathbf{S}^s) -\mathbf{A}_{\text{Fe}}\cdot\bm{\mathfrak{s}}^{s}\right], \label{eq:mean-field_fe}
\end{align}
where $\bar{s}$ is the complementary sublattice of $s$.
Eq.~\ref{eq:mean-field_fe} is the same as that in Ref.~\cite{Bamba2022} but Eq.~\ref{eq:mean-field-er} has an additional term coming from the inclusion of the $\text{Er}^{3+}$-anisotropy.

The equations of motion are linearized around the equilibrium  mean-fields, $\bar{\mathbf{B}}^s_{\text{Er/Fe}}\equiv \mathbf{B}_{\text{Er/Fe}}(\bar{\bm{\mathfrak{s}}}^\text{A/B},\bar{\mathbf{S}}^\text{A/B})$.
These equations of motion correspond to the effective Hamiltonians for the $\text{Er}$ and $\text{Fe}$ spins
\begin{align}
    \mathcal{H}^s_{\text{Er}} &= \mathfrak{g}_{\text{Er}}\mu_\text{B} \bm{\mathfrak{s}}^s\cdot \bar{\mathbf{B}}_{\text{Er}}^s = \mathfrak{g}_{\text{Er}} \mu_\text{B} \mathfrak{s}^s_{\parallel}|\bar{\mathbf{B}}^s_{\text{Er}}|\\
    \mathcal{H}^s_{\text{Fe}} &= \mathfrak{g}_{\text{Fe}}\mu_\text{B} \mathbf{S}^s\cdot \bar{\mathbf{B}}_{\text{Fe}}^s = \mathfrak{g}_{\text{Fe}} \mu_\text{B} \mathbf{S}^s_{\parallel}|\bar{\mathbf{B}}^s_{\text{Fe}}|,
\end{align}
where $\bm{\mathfrak{s}}_\parallel$ ($\mathbf{S}_\parallel$) is the spin operator in the direction parallel to the mean-field. 
The finite-temperature average of the spin operators then gives  the self-consistency equations
\begin{align}
    \braket{\mathfrak{s}^s_{\parallel}} &= -\tanh\left(\frac{\mathfrak{g}\mu_\text{B} |\bar{\mathbf{B}}^s_{\text{Er}}|}{2k_\text{B} T}\right)\\
    \braket{S^s_{\parallel}} &= -B_S \left(\frac{S \mathfrak{g}\mu_\text{B} |\bar{\mathbf{B}}^s_{\text{Fe}}|}{k_\text{B} T}\right),
\end{align}
where $B_S(x) = \frac{2S + 1}{2S} \coth \left (\frac{2S + 1}{2S} x \right ) - \frac{1}{2S} \coth \left ( \frac{x}{2S}  \right )$ is the Brillouin function.

The linearized equations of motion are thus found to be
\begin{align}
    \hbar \frac{d}{dt} \delta \bm{\mathfrak{s}}^s &= -\delta\bm{\mathfrak{s}}^s \times \mathfrak{g}\mu_\text{B} \mathbf{\bar{B}}_{\text{Er}}^s (\bm{\mathfrak{s}}^\text{A/B},\mathbf{S}^\text{A/B})  -\bar{\bm{\mathfrak{s}}}^s \times \mathfrak{g}\mu_\text{B} \mathbf{{B}}_{\text{Er}}^s (\delta\bm{\mathfrak{s}}^\text{A/B},\delta\mathbf{S}^\text{A/B}), \label{eq:lin-eom-er}\\
    \hbar \frac{d}{dt} \delta\mathbf{S}^s &= -\delta\mathbf{S}^s \times \mathfrak{g}\mu_\text{B} \mathbf{\bar{B}}_{\text{Er}}^s (\bm{\mathfrak{s}}^\text{A/B},\mathbf{S}^\text{A/B})-\bar{\mathbf{S}}^s \times \mathfrak{g}\mu_\text{B} \mathbf{{B}}_{\text{Er}}^s (\delta\bm{\mathfrak{s}}^\text{A/B},\delta\mathbf{S}^\text{A/B}), \label{eq:lin-eom-fe}
\end{align}
where $\delta\bm{\mathfrak{s}}^s$ and $\delta\mathbf{S}^s$ are the difference of the spin operators and their mean-field values. 

Eqs.~\ref{eq:lin-eom-er} and \ref{eq:lin-eom-fe} are solved to obtain the resonance frequencies.  As coupled homogeneous linear differential equations with constant coefficients they are solved by $\delta \bm{\mathfrak{s}}^s = \bm{\mathfrak{s}}^s(0) e^{i\omega^s_{\text{Er}}t}$ and $\delta \mathbf{S}^s =\delta \mathbf{S}^s(0) e^{i\omega^s_{\text{Fe}}t}$, where the frequencies $\omega$ and corresponding normal modes are  found by solving a set of linear equations, $i\omega\bm{s} = M\bm{s}$, where $\bm{s} = (\bm{\mathfrak{s}}^A,\mathfrak{s}^B,\mathbf{S}^A,\mathbf{S}^B)$ and $M$ is an anti-Hermitian matrix. 
The imaginary part of the eigenvalues of $M$ are the resonance frequencies.

\subsection*{\underline{Model fitting}}

\newcommand{\blank}{\,&}

Most model parameters are taken from previous literature, including all Fe-Er interaction parameters and most of the \ce{Fe^3+} subsystem. We fit other model parameters, summarized in Tab.~\ref{tab:modelParam}. Where previously parameters are available, our fits are consistent with prior work. 

We use the large majority of Ref.~\cite{Bamba2022}'s  model parameters for \ce{Fe^3+} subsystem, adjusting only $A_z$ and $\mathfrak{g}_x^{\text{Fe}}$. We also use that reference's $b$- and $c$-axis \ce{Er^3+}   $\mathfrak{g}$-factors along $b$ and $c$ axes. They are listed without uncertainty in Tab.~\ref{tab:modelParam}. We follow the notation and parametrization in \cite{Bamba2022}.

We adjust some model parameters that earlier experimental data was fairly  insensitive to -- mainly parameters of the \ce{Er^3+} subsystem, and introduce a new parameter, the anisotropy energy of the \ce{Er^3+} spins. We fit the model parameters to match the experimentally obtained THz and GHz magnetospectroscopy resonance positions by minimizing the cost function
\begin{equation}
    C=\sum_{i} \frac{w_i}{N_i} \sum_{j\in\text{$i$-th curve}} \frac{(f_{ij}  - \tilde{f}_{ij})^2}{\tilde{f}_{ij}^2},
\end{equation}
where $i$ indexes the dataset that is fit to (there are five sets: four spectral modes versus $H$ and the $H=0$ critical temperature, as described below), and $N_i$ is the number of experiment data points in the $i$-th curve. $w_i$ is the weight assigned to the $i$-th curve (described below), but fits are relatively insensitive to this choice; $j$ indices experiment data points in the $i$-th dataset; and  $f_{ij}$ and $\tilde{f}_{ij}$ are the calculated and measured observables frequencies of $i$-th mode at $j$-th data point, respectively.

The five datasets are the four spectral mode frequencies as a function of $H$, and the $H=0$ critical temperature. For the first four datasets, ${\tilde f}_{ij}$ is the resonance frequency of  mode $i$ and $f_{ij}$ is the mean field calculation of that mode at the $H$ and $T$ values corresponding to measurement $j$. 
The fifth ``dataset" is simply the temperature at which  $\omega_+$ has a cusp. We weigh most datasets with $0.2$;  the exceptions are the \ce{Fe^3+} mode, with its critical softening, we weigh slightly more, with weight 0.3, and the \ce{Er^+} slightly less, with weight 0.1.

 Figures\,S5A--B  and Fig.\, 3 of the main text  compares the mean-field calculations with the experimenta data they were fit to.  Figure\,S5C shows the predictions for $T=10$\,K using  these parameters (this data was not included in the fit). 
Tab.~\ref{tab:modelParam} presents the obtain model parameters as quantities with the quoted uncertainties using the method in \cite{Vugrin2007} without including experiment measurement errors.

\begin{table}
    \centering
    \begin{tabular}{c|c|c}
        \hline
        Fe$^{3+}$ subsystem & Er$^{3+}$ subsystem & Fe$^{3+}$--Er$^{3+}$ interaction \\
        \hline
         $J_\text{Fe} = 4.96\,\text{meV}$ & $J_\text{Er} = 0.01328(5) \,\text{meV}$ & $J = 0.6 \,\text{meV}$ \\
         $D^\text{Fe}_y =  -0.107 \,\text{meV}$ & $a_x = 0.124(4) \,\text{meV}$ & $D_x = 0.034 \,\text{meV}$ \\
        $A_x = 0.0073 \,\text{meV}$ & $a_z = 0.1480(3) \,\text{meV}$ & $D_y = 0.003 \,\text{meV}$ \\
        $A_z = 0.0176(3) \,\text{meV}$ & $a_{xz} = 0\,\text{meV}$ & \\
        $A_{xz} = 0\,\text{meV}$ & $\mathfrak{g}^\text{Er}_x = 4.16(8)$ & \\
        $\mathfrak{g}^\text{Fe}_x = 3.5734(3)$ & $\mathfrak{g}^\text{Er}_y = 3.4$ & \\
        $\mathfrak{g}^\text{Fe}_y = 2$ & $\mathfrak{g}^\text{Er}_z = 9.6$ & \\
        $\mathfrak{g}^\text{Fe}_z = 0.6$ & & \\
         \hline
    \end{tabular}
    \caption{Mean field model parameters from fitting and \cite{Bamba2022}. Listed parameter uncertainties are errors of fit, determined by the method in \cite{Vugrin2007}.}
    \label{tab:modelParam}
\end{table}


\subsection*{\underline{Extended Dicke model}}
The derivation of the extended Dicke model from the spin Hamiltonian follows ``Derivation of extended Dicke Hamiltonian" of Ref.~\cite{Bamba2022}.
We briefly outline this here.
The spin Hamiltonian is split in three parts, namely $\mathcal{H}_{\text{Fe}}$, $\mathcal{H}_{\text{Er}}$ and $\mathcal{H}_{\text{Er-Fe}}$.

First, we rewrite the $\text{Fe}^{3+}$ subsystem in terms of magnon annihilation and creation operators.
The magnons are collective spin-fluctuations above mean-field configuration for the bare $\text{Fe}^{3+}$ subsystem.
The  ground state is the same as previous studies \cite{Bamba2022,Li2018,Herrmann1963}, 
\begin{equation}
    \bar{\mathbf{S}}_0^\text{A} = \begin{pmatrix} 
     S \sin\beta_0 \\
     0 \\
     -S\cos\beta_0
    \end{pmatrix}, \hspace{0.4in}
    \bar{\mathbf{S}}_0^\text{B}=\begin{pmatrix}
        S\sin\beta_0 \\
        0 \\
        S \cos\beta_0
    \end{pmatrix}
\end{equation}
where $\beta_0$ is the mean-field canting angle of the $\text{Fe}^{3+}$-spins from the $a$ axis \cite{Herrmann1963,Bamba2022}
\begin{equation}
    \beta_0 = -\frac{1}{2}\arctan\left[\frac{z_{\text{Fe}}D^y_{\text{Fe}}}{z_{\text{Fe}}J_{\text{Fe}} - A^x_{\text{Fe}} + A^z_{\text{Fe}}}\right].
\end{equation}
We neglect the non-diagonal $\text{Fe}^{3+}$-anisotropy term, $A_{\text{Fe}}^{xz}$. 

Spin-wave theory  yields\cite{Herrmann1963,Bamba2022} 
\begin{equation}
    \mathcal{H}_{\text{Fe}} \approx \sum_{k=0,\pi} \hbar \omega_k a_k^\dagger a_k + \text{const},
\end{equation}
with $\omega_k = \mathfrak{g}_{\text{Fe}}\mu_\text{B}/\hbar\sqrt{(b\cos k - a)(d\cos k + c)}$ and
\begin{align}
    a &= [S/(\mathfrak{g}_{\text{Fe}}^x\mu_\text{B})]\left[-A^z_\text{Fe} - A^x_\text{Fe} - (z_\text{Fe}J_\text{Fe} + A^z_\text{Fe}-A^x_\text{Fe})\cos(2\beta_0) + z_\text{Fe}D^y_\text{Fe}\sin(2\beta_0)\right]\\
    b &= [S/(\mathfrak{g}_{\text{Fe}}^x\mu_\text{B})]z_\text{Fe}J_\text{Fe}\\
    c &= [S/(\mathfrak{g}_{\text{Fe}}^x\mu_\text{B})]\left[(z_\text{Fe}J_\text{Fe} + 2 A^z_\text{Fe} - 2A^x_\text{Fe})\cos(2\beta_0) + z_\text{Fe}D^y_\text{Fe}\sin(2\beta_0)\right] \\
    d &= [S/(\mathfrak{g}_{\text{Fe}}^x\mu_\text{B})]\left[-z_\text{Fe}J_\text{Fe}\cos(2\beta_0) -z_\text{Fe}D^y_\text{Fe}\sin(2\beta_0)\right].
\end{align}
The derivation is as is from Ref.~\cite{Bamba2022}.
First, we compute the Heisenberg equations of motion of $\mathbf{S}$ for $\mathcal{H}_{\text{Fe}}$.
Then we linearize them by expressing $\mathbf{S}_i^{\text{A/B}} = \bar{\mathbf{S}}_0^{\text{A/B}} + \delta \mathbf{S}_i^{\text{A/B}}$.
The effective Hamiltonian of the linearized equations of motion are decoupled harmonic oscillators.
These harmonic oscillators are the magnons.
The spin fluctuations in terms of ladder operators of the harmonic oscillators are
\begin{align}
    \delta\mathbf{S}_i^\text{A} \approx \sqrt{\frac{S}{2N_0}}\begin{pmatrix}
        -(T_0 - T_\pi)\cos\beta_0 \\
        (Y_0 - Y_\pi) \\
        -(T_0 - T_\pi)\sin\beta_0 
    \end{pmatrix}, \label{eq:spin-fluc-A} \\
    \delta\mathbf{S}_i^\text{B} \approx \sqrt{\frac{S}{2N_0}}\begin{pmatrix}
        (T_0 + T_\pi)\cos\beta_0 \\
        (Y_0 + Y_\pi) \\
        -(T_0 + T_\pi)\sin\beta_0 
    \end{pmatrix}, \label{eq:spin-fluc-B}
\end{align}
where 
\begin{align}
    T_k &= \left(\frac{b\cos k - a}{d\cos k - a}\right)^{1/4} \frac{a^\dagger_{-k} + a_k}{\sqrt{2}} \\
    Y_k &= \left(\frac{d\cos k + c}{b\cos k - a}\right)^{1/4} \frac{i(a^\dagger_{-k} - a_k)}{\sqrt{2}}.
\end{align}
We only keep $k= 0,\pi$ contributions, the  the qFM and the qAFM modes of $\text{Fe}^{3+}$ respectively, as there is negligible coupling to the other modes.
Henceforth these modes are labeled as $a_{\text{qFM}}$ and $a_{\text{qAFM}}$.

Next, we apply the collective spin approximation on the $\text{Er}^{3+}$-spins,
\begin{align}
    \bm{\mathfrak{s}}_i^s &= \frac{1}{N_0}\sum_{i=1}^{N_0}\bm{\mathfrak{s}}_i^s \\
    &\equiv  \frac{1}{N_0}\bm{\Sigma}_i^s,
\end{align}
finding
\begin{align}
    \mathcal{H}_{\text{Er}} &\approx \mathfrak{g}_{\text{Er}}^x \mu_\text{B} \Sigma_x^+ B_x^{\text{DC}} + z_{\text{Er}} J_{\text{Er}} \sum_{i=1}^{N_0} \bm{\mathfrak{s}}^\text{A}_i \cdot \sum_{i'=1}^{N_0} \frac{\bm{\mathfrak{s}}^\text{B}_{i'}}{N_0} - \sum_i \sum_{\xi=x,z}A^\xi_{\text{Er}}\sum_{s=\text{A,B}} \left(\frac{1}{N_0}\sum_j \mathfrak{s}_{j,\xi}^s\right)^2\\
    &=\mathfrak{g}_{\text{Er}}^x \mu_\text{B} \Sigma_x^+ B_x^{\text{DC}} + \frac{z_{\text{Er}} J_{\text{Er}}}{N_0} \bm{\Sigma}^\text{A}\cdot\bm{\Sigma}^\text{B} - \sum_{s=\text{A,B}}\sum_{\xi=x,z} \frac{A^\xi_{\text{Er}}}{N_0} (\Sigma^s_\xi)^2,
\end{align}
where $\Sigma_\xi^{\pm} = \Sigma_\xi^\text{A} \pm \Sigma_\xi^\text{B}$.

Finally, we rewrite $\mathcal{H}_{\text{Er-Fe}}$ in terms of the spin-fluctuation operators of the $\text{Fe}^{3+}$-spins and the collective spin operators of $\text{Er}^{3+}$-spins to give
\begin{align}
    \mathcal{H}_{\text{Er-Fe}} &= 4S(J\sin\beta_0 + D_y\cos\beta_0) \Sigma_x^+ + (-4SD_x\cos\beta_0)\Sigma_y^-+ \sqrt{\frac{S}{N_0}}[(J\cos\beta_0 - D_y\sin\beta_0)T_\pi\Sigma_x^+   \\
    &+ JY_0\Sigma_y^++ (D_x\sin\beta_0)T_\pi\Sigma_y^-+ D_x Y_\pi\Sigma_z^- - (J\sin\beta_0 +D_y\cos\beta_0)T_0\Sigma_z^+ ].
\end{align}
We have different pre-factors of the parameters compared to Ref.~\cite{Bamba2022} since they have extra factors of $1/2$ as the $\text{Er}^{3+}$-spins are modeled as $1/2 \bm{\sigma}^s$, whereas we absorb this in our spin variables $\bm{\mathfrak{s}}^s$.

The complete spin-Dicke Hamiltonian, expressed in terms of the magnonic operators and collective $\text{Er}^{3+}$-spin operators is
\begin{equation}
\begin{aligned}
    \mathcal{H}_\text{Dicke} \approx &\sum_{\text{m}:\{\text{qFM},\text{qAFM}\}} \hbar\omega_{\text{m}} a_{\text{m}}^\dagger a_{\text{m}} + E_x \Sigma_x^+ + E_y\Sigma_y^- + \mu_\text{B}\mathfrak{g}_\text{Er}^x B_x^\text{DC}\Sigma_x^+ + \frac{z_\text{Er}J_\text{Er}}{N_0}\bm{\Sigma}^\text{A}\cdot\bm{\Sigma}^\text{B} \\
     &-\sum_{\xi=x,z}\sum_{s=A,B}\frac{A^\xi_\text{Er}}{N_0}(\Sigma_\xi^\text{s})^2 + \frac{g_x}{\sqrt{N_0}} \left(a_{\text{qAFM}}^\dagger + a_{\text{qAFM}}\right)\Sigma_x^+ + \frac{ig_y}{\sqrt{N_0}}\left(a_{\text{qFM}}^\dagger - a_{\text{qFM}}\right)\Sigma_y^+\\   
     & + \frac{g_{y'}}{\sqrt{N_0}}\left(a_{\text{qAFM}}^\dagger + a_{\text{qAFM}}\right)\Sigma_y^- + \frac{ig_z}{\sqrt{N_0}}\left(a_{\text{qAFM}}^\dagger - a_{\text{qAFM}}\right)\Sigma_z^- + \frac{g_{z'}}{\sqrt{N_0}}\left(a_{\text{qFM}}^\dagger + a_{\text{qFM}}\right)\Sigma_z^+. 
    \end{aligned}
\end{equation}
 $E_x = 4S(J\sin\beta_0 + D_y\cos\beta_0)$, $E_y = -4SD\cos\beta_0$ and the five coupling strengths are 
\begin{align}
    g_x &= \sqrt{ x S} \left(J \cos \beta_0 - D_y\sin \beta_0\right) \left(\frac{b+a}{d-c}\right)^{1/4} \\
    g_y &= \sqrt{ x S} J \left(\frac{d+c}{b-a}\right)^{1/4} \\
    g_{y'} &= \sqrt{ x S} D_x\sin \beta_0 \left(\frac{b+a}{d-c}\right)^{1/4}\\
    g_{z} &= \sqrt{xS} D_x \left(\frac{d-c}{b+a}\right)^{1/4}\\
    g_{z'} &= \sqrt{ x S}\left(-J\sin \beta_0 - D_y \cos \beta_0\right) \left(\frac{b-a}{d+c}\right)^{1/4}.
\end{align}
$x = \tanh\left[\frac{|E_x + \mu_\text{B}\mathfrak{g}_{\text{Er}}^xB_x^{\text{DC}}|}{2k_\text{B} T}\right]$ is the spin-dilution factor that incorporates the effect of temperature.


\clearpage
\begin{figure}[t]
		\centering
		\includegraphics[width=1\linewidth]{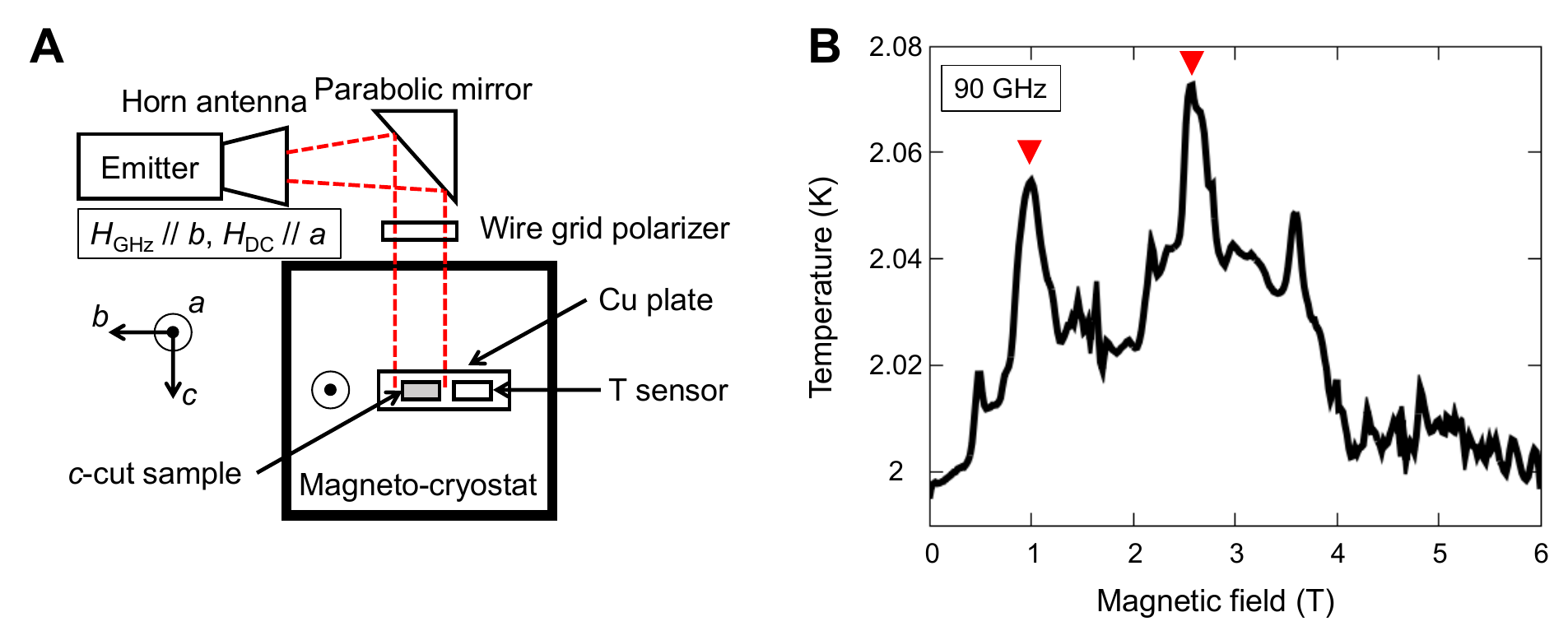}
	\end{figure}
 
\noindent {Figure S1: \textbf{Thermal detection of magnetic resonances of Er$^{3+}$ spins.}} (\textbf{A})~A schematic of the setup for thermal detection. (\textbf{B})~The sample temperature as a function of the static magnetic field with 90~GHz illumination. When the incident photon energy coincides with the transition energy of magnetic resonances, the temperature increases.

\clearpage
\begin{figure}[t]
		\centering
		\includegraphics[width=1\linewidth]{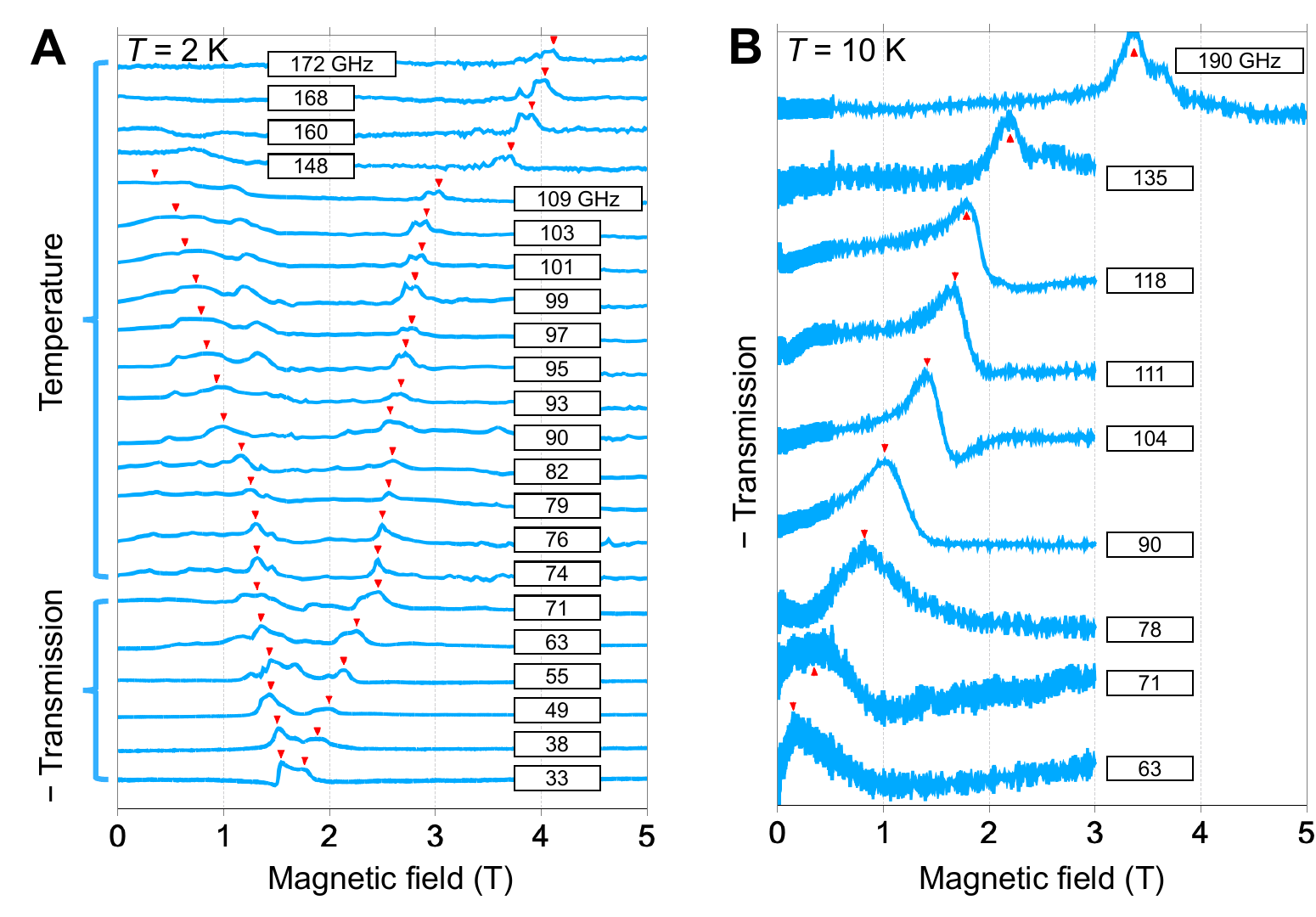}
	\end{figure}
 
\noindent {Figure S2: \textbf{Raw data of GHz measurements with 0-to-1 scaled.}} (\textbf{A})~From 33 to 71~GHz (74 to 172~GHz), transmission (temperature) spectra as a function of the magnetic field at 2~K. These data are used to generate the two middle panels and the bottom panel in Fig.\,2C. (\textbf{B})~From 63 to 190~GHz, transmission spectra as a function of the magnetic field at 10~K. Red triangles indicate the resonance peak positions that are used in Fig.\,3B (red circles).

\clearpage
\begin{figure}[t]
		\centering
		\includegraphics[width=1\linewidth]{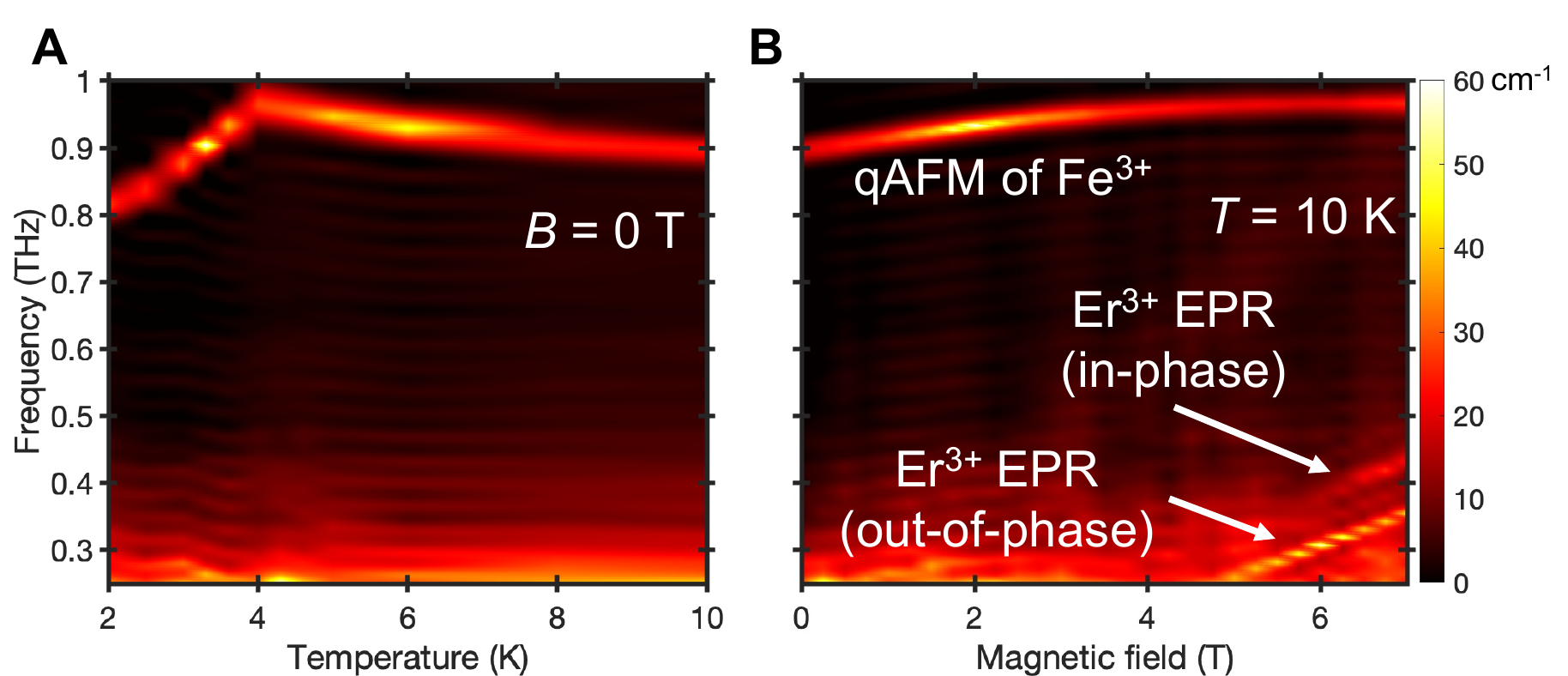}
	\end{figure}
 
\noindent {Figure S3: \textbf{Terahertz absorption coefficient spectra}} (\textbf{A})~Absorption coefficient spectra as a function of temperature in THz-TDS, showing a kink at the phase boundary. (\textbf{B})~Absorption coefficient spectra as a function of magnetic field in THz-TDMS. The qAFM mode of Fe$^{3+}$ and two Er$^{3+}$ EPR modes are observed. The out-of-phase mode is plotted in Fig.\,3B (right panel). All three modes are reproduced by our mean-field calculation in Fig.\,S5C.

\clearpage
\begin{figure}[t]
		\centering
		\includegraphics[width=1\linewidth]{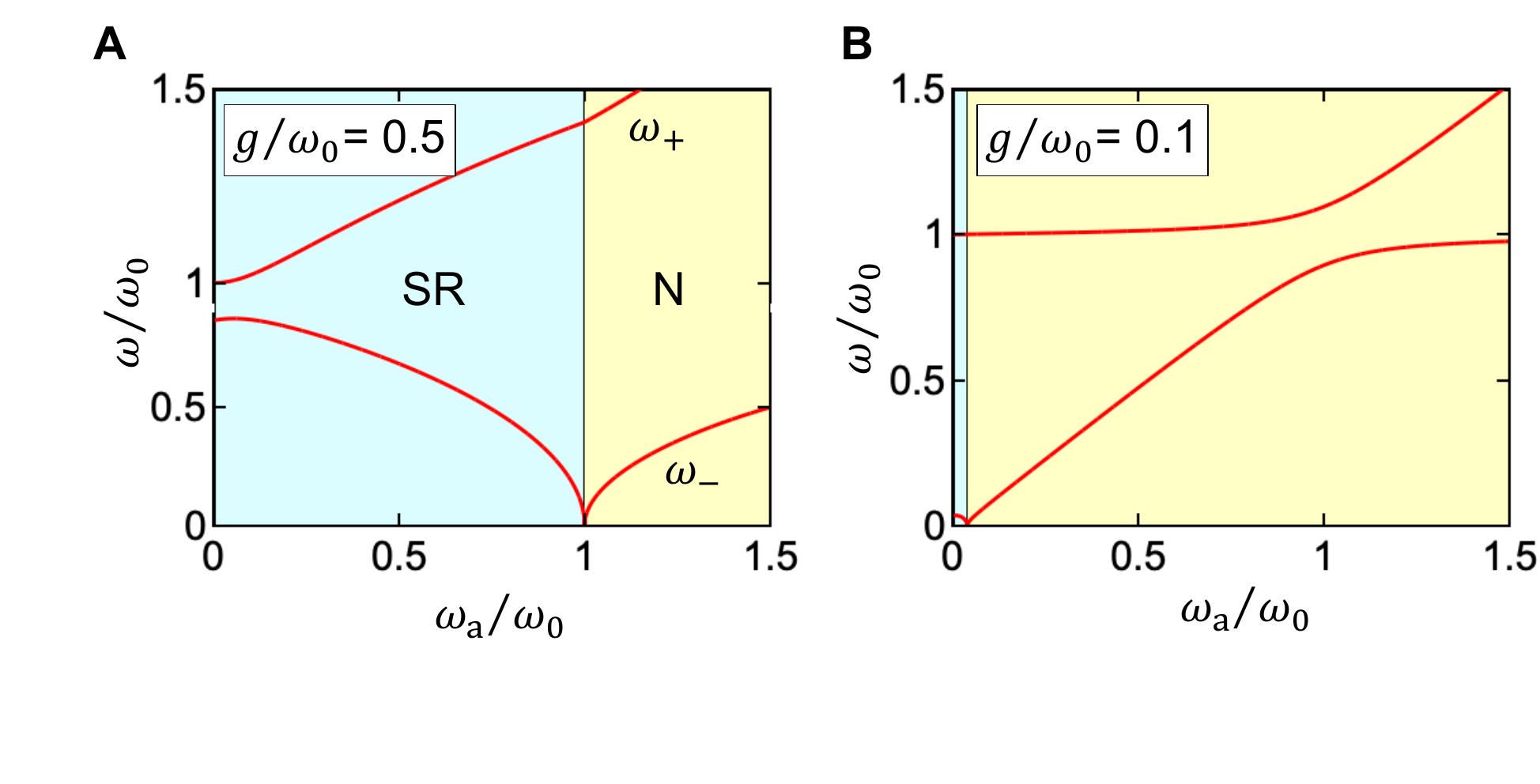}
	\end{figure}
 
\noindent {Figure S4: \textbf{Occurrence of the superradiant phase transition at finite detuning}} (\textbf{A})~Normalized frequencies of the upper-polariton ($\omega_{+}$) and lower-polariton ($\omega_{-}$) modes as a function of $\omega_\text{a}/\omega_\text{0}$ calculated using the Dicke model without the $A^2$ term with $g/\omega_\text{0} = 0.5$ and (\textbf{B}) with $g/\omega_\text{0} = 0.1$. 

\clearpage
\begin{figure}[t]
		\centering
		\includegraphics[width=1\linewidth]{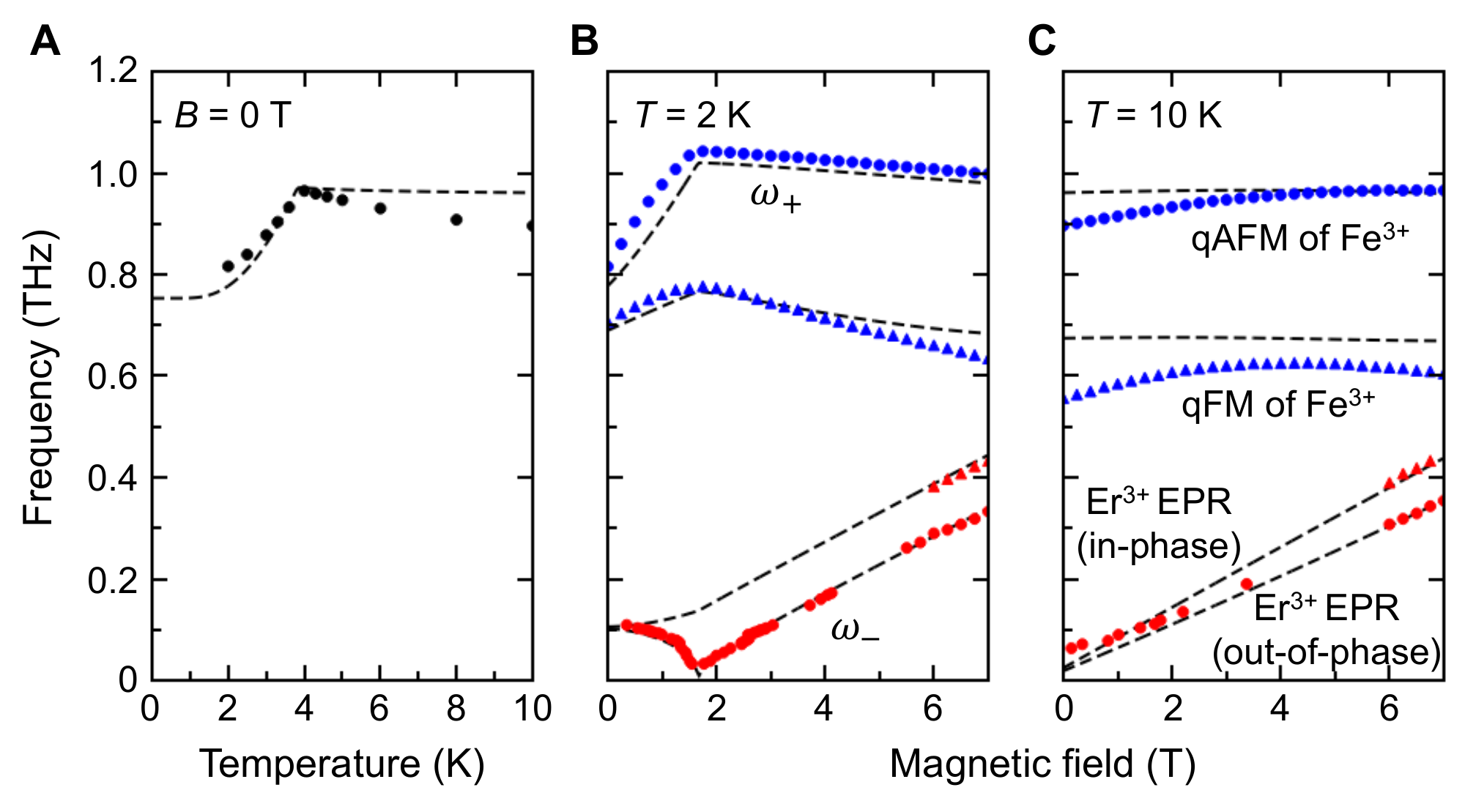}
	\end{figure}
 
\noindent {Figure S5: \textbf{Comparison between experimental data and fitting results for the extended data}} (\textbf{A})~Temperature dependent absorption peaks of Fe$^{3+}$ ions. (\textbf{B})~Magnetic field dependent absorption peaks of all four modes at 2~K, and (\textbf{C}) at 10~K. We present our fitting curves in A and B, and calculations for 10~K in C. The blue circles, red triangles, and red circles in B are extracted from Fig.\,2B, while those in C are from Fig.\,S3B. The blue triangles correspond to qFM of Fe$^{3+}$, obtained from separate experiments with a 90\textdegree~rotated incident THz magnetic field polarization.

\clearpage
\begin{figure}[t]
		\centering
		\includegraphics[width=0.5\linewidth]{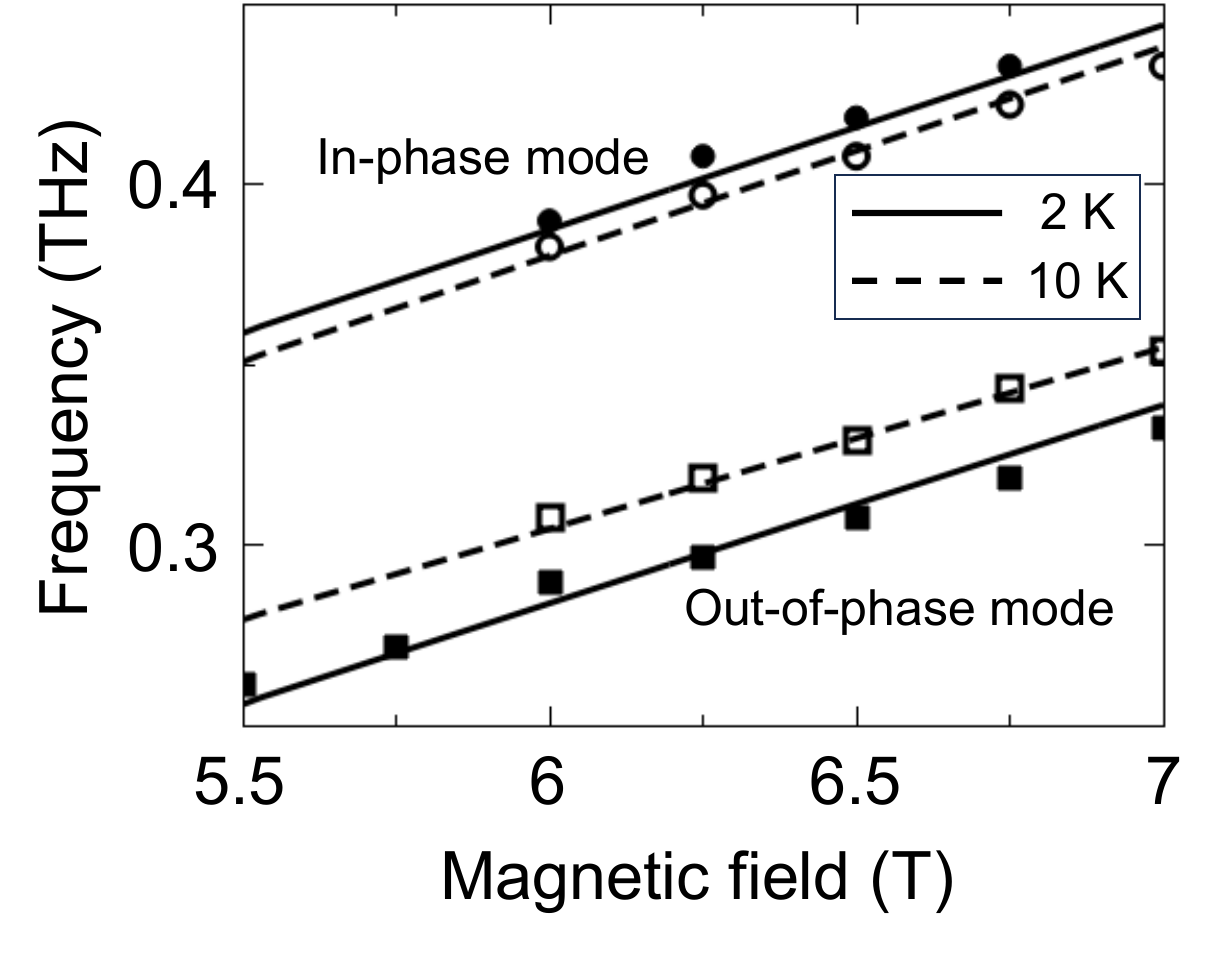}
	\end{figure}
 
\noindent {Figure S6: In-phase and out-of-phase modes of Er$^{3+}$ spins below and above the critical temperature, showing shifts only for the 2~K case. Lines (theory), points (experiment)}.

\clearpage
\begin{figure}[t]
		\centering
		\includegraphics[width=0.5\linewidth]{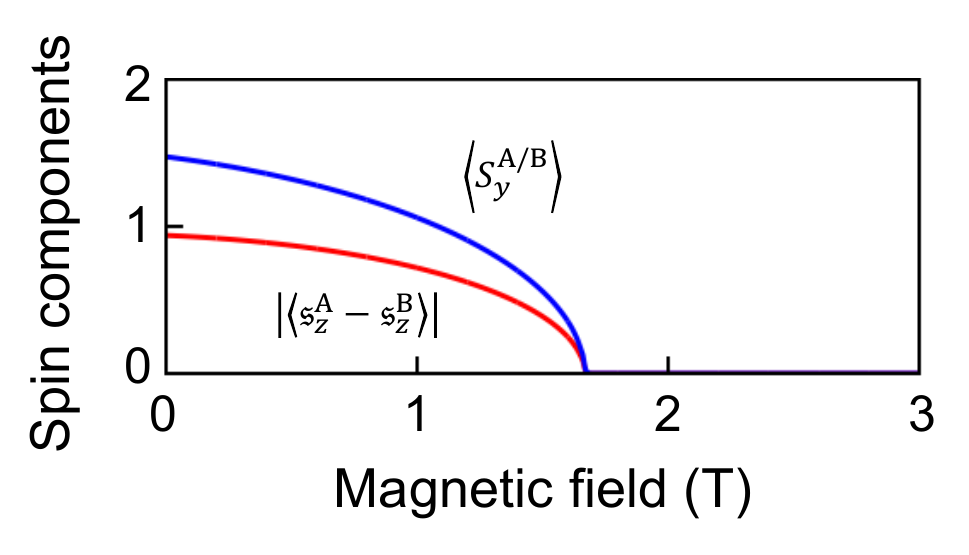}
	\end{figure}
 
\noindent {Figure S7: Two order parameters evidencing the magnonic SRPT calculated at 2~K.}

\end{document}